\documentclass{elsart5p}
\usepackage{natbib}
\usepackage{xspace}

\usepackage{epsfig}

\usepackage{amssymb}

\newlength{\hoehe}

\def\cref#1{Chapt.\,\ref{#1}}
\def\Cref#1{Chapter~\ref{#1}}
\def\sref#1{Sect.\,\ref{#1}}

\def\fref#1{Fig.\,\ref{#1}}
\def\ffref#1{Figs.\,\ref{#1}}

\def\eref#1{(\ref{#1})}

\def\lleft{\textit{left}}
\def\rright{\textit{right}}

\def\TTop{\textit{Top}}
\def\BBottom{\textit{Bottom}}

\def\1{\footnotemark[1]}
\def\and{\& }
\def\Cerenkov{\v{C}erenkov\xspace}

\def\etal{et al.\xspace}
\def\gcm2{g/cm$^2$\xspace}
\def\li{$\lambda_i$\xspace}
\def\lnA{\langle\ln A\rangle}
\def\lg{{\rm lg}\xspace}

\def\modell{poly-gonato model\xspace}

\def\xn{$X_0$\xspace}
\def\Xmax{$X_{max}$\xspace}

\def\ga{\raisebox{-0.2em}{\,$\stackrel{\scriptstyle>}{\scriptstyle\sim}$\,}}

\def\line{---}
\def\dashed{-\,-\,-}
\def\dotted{$\cdot\cdot\cdot$}
\def\dashdot{-$\cdot$-$\cdot$}

\begin{document}

\begin{frontmatter}

% Title, authors and addresses

% use the thanksref command within \title, \author or \address for footnotes;
% use the corauthref command within \author for corresponding author footnotes;
% use the ead command for the email address,
% and the form \ead[url] for the home page:
% \title{Title\thanksref{label1}}
% \thanks[label1]{}
% \author{Name\corauthref{cor1}\thanksref{label2}}
% \ead{email address}
% \ead[url]{home page}
% \thanks[label2]{}
% \corauth[cor1]{}
% \address{Address\thanksref{label3}}
% \thanks[label3]{}

\title{Cosmic-ray composition and its relation to shock acceleration by
       supernova remnants\thanksref{label1}}
 \thanks[label1]{Invited talk given at the 36th COSPAR Scientific Assembly
                 Beijing, China, 16 -- 23 July 2006.}

% use optional labels to link authors explicitly to addresses:
% \author[label1,label2]{}
% \address[label1]{}
% \address[label2]{}

\author{J\"org R. H\"orandel}

\address{University of Karlsruhe, Institute for Experimental Nuclear Physics,
P.O. Box 3640, 76021 Karlsruhe, Germany}
\ead[url]{www-ik.fzk.de/$\sim$joerg}

\begin{abstract}
An overview is given on the present status of the understanding of the origin
of galactic cosmic rays. Recent measurements of charged cosmic rays and photons
are reviewed. Their impact on the contemporary knowledge about the sources and
acceleration mechanisms of cosmic rays and their propagation through the Galaxy
is discussed. Possible reasons for the knee in the energy spectrum and
scenarios for the end of the galactic cosmic-ray component are described.
\end{abstract}


\end{frontmatter}

% main text
\section{Introduction}

The origin of high-energy cosmic rays is one of the open questions in
astroparticle physics. The fully ionized atomic nuclei reach the Earth from
outside the solar system with energies from the GeV range up to $10^{20}$~eV.
Most of them are assumed to originate in the Milky Way.  At the highest
energies, exceeding $10^{17}$~eV, the particles are usually considered of
extragalactic origin. This review focuses on galactic cosmic rays.
To distinguish between different models of the cosmic-ray origin requires
detailed measurements of the energy spectrum, mass composition, and arrival
direction of charged cosmic rays. Additional and complementary information is
obtained through the measurements of high-energy photons up to TeV energies.

This review starts with a short overview on detection methods typically applied
to measure the composition of cosmic rays on satellites, balloons, and at 
ground level (\sref{expsect}).
The propagation of cosmic rays through the Galaxy is in the focus of
\sref{propsect}.
Recent progress concerning the sources and acceleration of the high-energy
particles is described in \sref{sourcesect}.
One of the headstones in understanding the origin of galactic cosmic rays is to
know the reasons for the knee in the energy spectrum. In \sref{kneesect} recent
measurements of the all-particle energy spectrum, the mean mass, and spectra
for individual elements are compiled and their impact on the contemporary
knowledge of the origin of galactic cosmic rays is discussed.
The end of the galactic cosmic-ray spectrum and the transition to an
extragalactic component is briefly illuminated in \sref{extragsect}.

\section{Experimental Techniques} \label{expsect}

\begin{figure}[t] \centering
 \epsfig{file=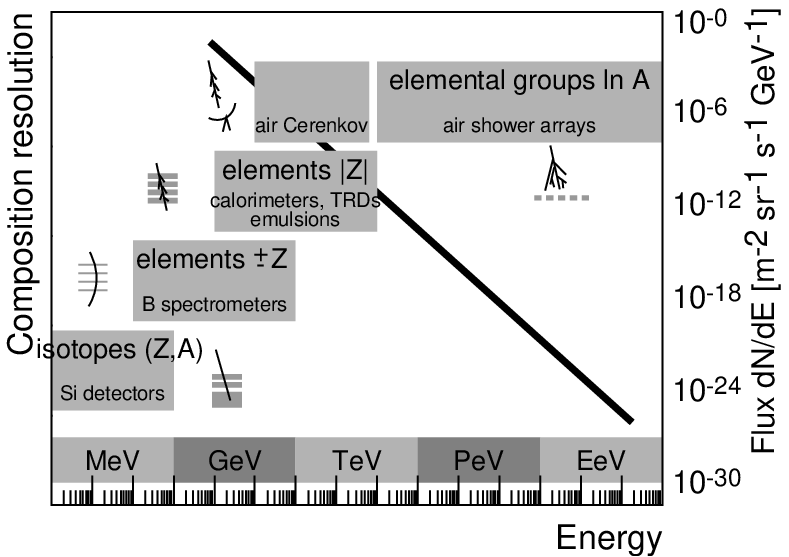,width=\columnwidth}
 \caption{Illustrative sketch of the composition resolution achieved by
	  different cosmic-ray detection techniques as function of energy.
	  Over the energy range shown the flux of cosmic rays decreases by
	  about 30 orders of magnitude as indicated on the right-hand scale.}
 \label{massres}	  
\end{figure}

The energies of cosmic rays extend from the GeV domain up to $10^{20}$~eV.
Within this range the particle flux decreases by about 30 orders of magnitude.
This has implications on the accuracy to determine the mass composition of
cosmic rays. At low energies the flux is large enough to build sophisticated
detectors with an active area of a few 100~cm$^2$ to measure the abundance of
individual isotopes. On the other hand, at the highest energies, where only a
few particles are expected per km$^2$ and century, huge ground based
installations are necessary to measure secondary products generated by cosmic
rays in the atmosphere and the (average) mass can be estimated coarsely
only. The situation is sketched figuratively in \fref{massres}.

At energies in the MeV range sophisticated silicon detectors, operated in outer
space, like the Ulysses HET \citep{ulysses} or the ACE/CRIS (0.025~m$^2$\,sr)
\citep{cris} experiments can identify individual isotopes, fully characterized
by simultaneous measurements of their energy, charge, and mass $(E,Z,A)$.
Since the particles have to be absorbed completely in a silicon detector this
technique works up to energies of a few GeV only.

In the GeV domain particles are registered with magnetic spectrometers on
stratospheric balloons, like the BESS instrument \citep{bess}. They are the
only detectors discussed here which are able to identify the sign of the
particles charge. All other detectors rely on the fact that the specific
ionization loss is $dE/dx\propto Z^2$, thus $|Z|$ is derived from the
measurements.  Magnet spectrometers are the only detectors suitable to
distinguish between matter and antimatter as e.g.  $e^+-e^-$, $p-\bar{p}$, or
$\mbox{He}-\overline{\mbox{He}}$.  The particle momentum is derived from the
curvature of the trajectory in a magnetic field, which limits the usage of
these detectors to energies approaching the TeV scale.

At higher energies particles are measured with balloon borne instruments on
circumpolar long duration flights \citep{ldb}. Individual elements are
identified, characterized by their charge and energy. $|Z|$ is determined
through $dE/dx$ measurements. Experimentally most challenging is the energy
measurement. In calorimeters the particles need to be (at least partly)
absorbed. The weight of a detector with a thickness of one hadronic interaction
length (\li) and an area of 1~m$^2$ amounts to about 1~t.  Due to weight
limitations actual detectors like ATIC \citep{atic} or CREAM \citep{creamexp}
have to find an optimum between detector aperture and energy resolution,
resulting in relatively thin detectors with a thickness of 1.7~\li (ATIC) or
0.7~\li (CREAM) only.  The measurement of transition radiation from cosmic-ray
particles allows to build large detectors with reasonable weight. The largest
cosmic-ray detector ever flown on a balloon, the TRACER experiment
\citep{gahbauer} has an aperture of 5~m$^2$\,sr. During a long duration balloon
flight over Antarctica \citep{jrhcospar} and another flight from Sweden to
Alaska \citep{boylecospar}, up to now an exposure of 70~m$^2$\,sr\,d has been
reached with this experiment. With such exposures the energy spectra for
individual elements can be extended to energies exceeding $10^{14}$~eV.

To access higher energies very large exposures are necessary. At present
reached only in ground based experiments, registering extensive air showers.
In the TeV regime (small) air showers are observed with imaging \Cerenkov
telescopes such as the HEGRA \citep{hegrap}, HESS \citep{hessexp}, MAGIC
\citep{magic}, or VERITAS \citep{veritas} experiments. These instruments image
the trajectory of an air shower in the sky with large mirrors onto a segmented
camera.  

For primary particles with energies exceeding $10^{14}$~eV the particle
cascades generated in the atmosphere are large enough to reach ground level,
where the debris of the cascade is registered in large arrays of particle
detectors.  Two types of experiments may be distinguished: installations
measuring the longitudinal development of showers (or the depth of the shower
maximum) in the atmosphere and apparatus measuring the density (and energy) of
secondary particles at ground level.

The depth of the shower maximum is measured in two ways.  Light-integrating
\Cerenkov detectors like the BLANCA \citep{blanca} or TUNKA \citep{tunka}
experiments are in principle arrays of photomultiplier tubes with light
collection cones looking upwards in the night sky, measuring the lateral
distribution of \Cerenkov light at ground level. The depth of the shower
maximum and the shower energy is derived from these observations.  Imaging
telescopes as in the HiRes \citep{hiresxmax} or AUGER \citep{augerexp}
experiments observe an image of the shower on the sky through measurement of
fluorescence light, emitted by nitrogen molecules, which had been excited by
air shower particles. These experiments rely on the fact that the depth of the
shower maximum for a primary particle with mass $A$ relates to the depth of the
maximum for proton induced showers as 
\begin{equation} \label{xmaxeq}
 X_{max}^A=X_{max}^p-X_0\ln A, 
\end{equation}
where $X_0=36.7$~\gcm2 is the radiation length in air
\citep{matthewsheitler,jrherice06}.

An example for an air shower array is the KASCADE-Grande experiment
\citep{kascadenim,grande}, covering an area of 0.5~km$^2$. The basic idea is to
measure the electromagnetic component in an array of unshielded scintillation
detectors and the muons in scintillation counters shielded by a lead and iron
absorber, while the hadronic component is measured in a large calorimeter
\citep{kalonim}. The total number of particles at observation level is obtained
through the measurement of particle densities and the integration of the
lateral density distribution \citep{kascadelateral}.  The direction of air
showers is reconstructed through the measurement of the arrival time of the
shower particles in the individual detectors. Measuring the electron-to-muon
ratio in showers, the mass of the primary can be inferred. A Heitler model of
hadronic showers \citep{jrherice06} yields the relation
\begin{eqnarray} \label{emratioeq}
 \frac{N_e}{N_\mu}\approx35.1
     \left(\frac{E_0}{A\cdot1~\mbox{PeV}}\right)^{0.15} \\
 \mbox{~or~} \lg\left(\frac{N_e}{N_\mu}\right)=C-0.065\ln A . \nonumber
\end{eqnarray}

This illustrates the sensitivity of air shower experiments to $\ln A$.  To
measure the composition with a resolution of 1 unit in $\ln A$ the shower
maximum has to be measured to an accuracy of about 37~\gcm2 (see \eref{xmaxeq})
or the $N_e/N_\mu$ ratio has to be determined with an relative error around
16\% (see \eref{emratioeq}). Due to the large intrinsic fluctuations in air
showers, with existing experiments at most groups of elements can be
reconstructed with $\Delta\ln A\approx0.8-1$.

\begin{figure}[t] \centering
 \epsfig{file=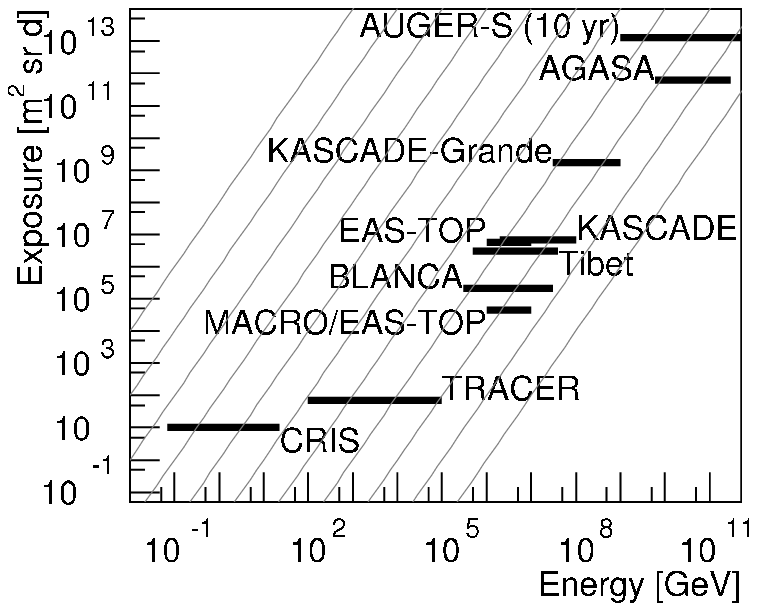,width=\columnwidth}
 \caption{Exposure of cosmic-ray experiments as function of energy for
          CRIS \citep{cris-time},
	  TRACER antarctic and Sweden LDB flights
	  \citep{jrhcospar,boylecospar},
	  MACRO/EAS-TOP \citep{eastop-macro-lna},
	  BLANCA \citep{blanca},
	  Tibet \citep{tibetasg03},
	  EAS-TOP \citep{eastope},
	  KASCADE \citep{ulrichapp},
	  KASCADE-Grande (estimated 3~yr) \citep{grande},
	  AGASA, and
	  AUGER south estimated 10~yr \citep{augerexp}.
	  The grey lines are $\propto E^{-2}$.}
 \label{exposure}  
\end{figure}

The exposure achieved by several experiments is shown in \fref{exposure}.  The
integral cosmic-ray flux roughly depends on energy as $\propto E^{-2}$.  Hence,
the grey lines ($\propto E^{-2}$) serve to compare the exposure of detectors in
different energy regimes. It can be recognized that the different experiments,
despite of their huge differences in physical size, ranging from the
$\approx300$~cm$^2$ CRIS space experiment to the $\approx3600$~km$^2$ AUGER air
shower experiment, have about equal effective sizes when the decreasing
cosmic-ray flux is taken into account.

\section{Propagation} \label{propsect}

\begin{figure*}[t] \centering
 \epsfig{file=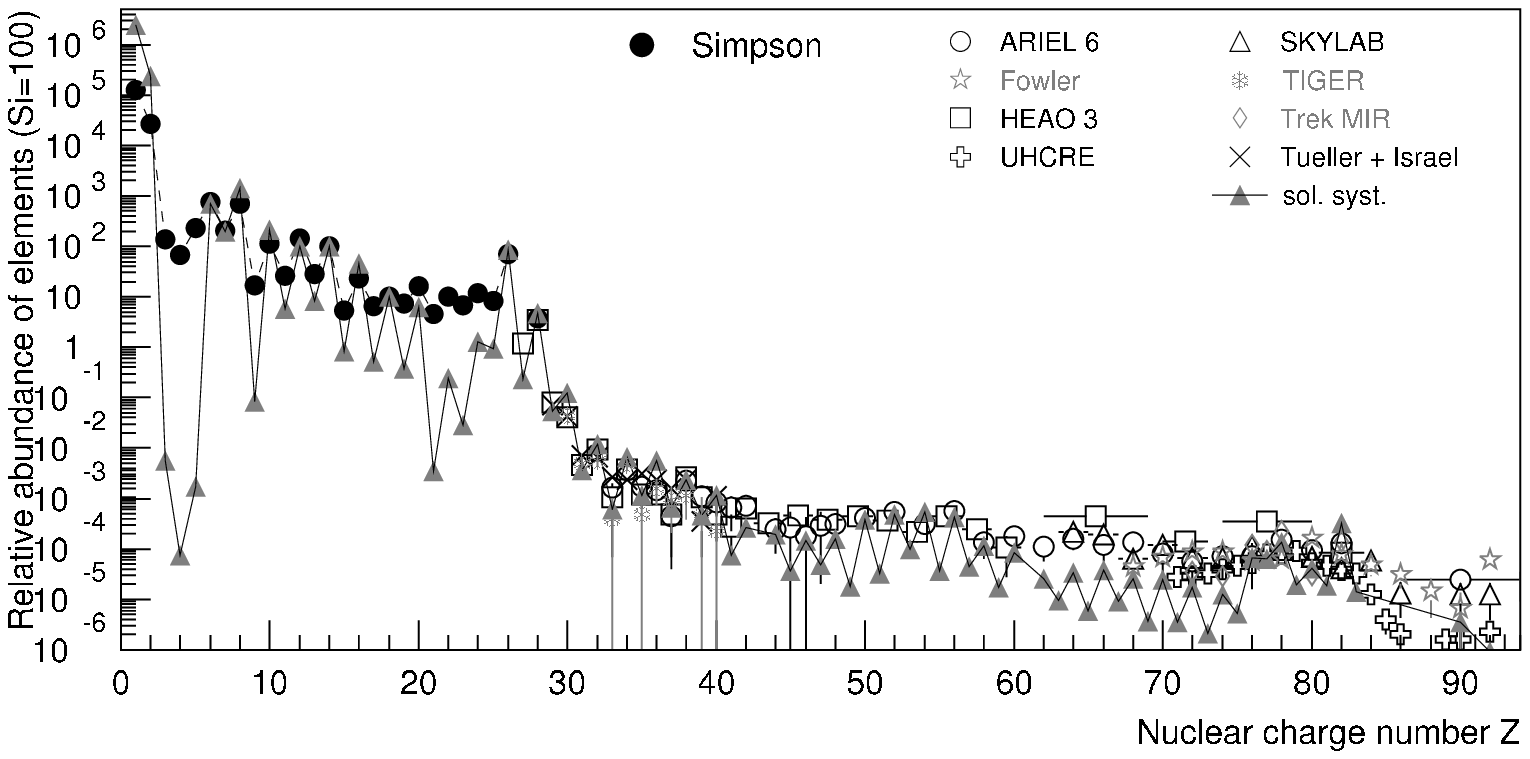,width=\textwidth}
 \caption{Abundance of elements in cosmic rays as function of their nuclear
          charge number $Z$ at energies around 1~GeV/n, 
          normalized to $Si=100$.
	  Abundance for nuclei with $Z\le28$ according to \citet{simpson}.
	  Heavy nuclei as measured by 
          ARIEL~6 \citep{ariel},
          \citet{fowler},
          HEAO 3 \citep{heao}, 
          SKYLAB \citep{skylab},
	  TIGER \citep{tiger},
          TREK/MIR \citep{trek}, 
          \citet{tueller}, as well as
          UHCRE \citep{uhcre}.
          In addition, the abundance of elements in the solar system is
	  shown according to \citet{lodders}.}
          
 \label{simpson}
\end{figure*}

The propagation of cosmic rays in the Galaxy can be described quantitatively,
taking into account energy loss, nuclear interactions, radioactive decay,
spallation production, and production by radioactive decay, using the equation
\citep{simpson}
\begin{eqnarray} \label{transporteq}
 \frac{\partial J_i}{\partial x} =
       \frac{\partial}{\partial E}\left(\frac{dE}{dx} J_i\right)
       -\frac{N_A \sigma_i^t J_i}{\bar{A}}
       -\frac{J_i}{\gamma\beta \rho c T_i} \nonumber \\
       +\sum_{j\ne i}\frac{N_A \sigma_{ij} J_i}{\bar{A}} 
       +\sum_{j\ne i} \frac{J_i}{\gamma\beta \rho c T_{ij}} \quad .
\end{eqnarray}
$J_i(x)$ is the flux of species $i$ after propagating through an amount of
matter $x$ $[$\gcm2$]$ subject to the reduction of $J_i(0)$, that represents
the source term. The other parameters are: $dE/dx$, the rate of ionization
energy loss; $\sigma_i^t$, the total inelastic cross section for species $i$;
$T_i$, the mean lifetime at rest for radioactive decay for species $i$;
$\sigma_{ij}$, the cross section for production of species $i$ from
fragmentation of species $j$; $T_{ij}$, the mean lifetime at rest for decay of
species $j$ into species $i$; $N_A$, Avogadro's Number; $\bar{A}$, the mean
atomic weight of the interstellar gas; and $\gamma$, the Lorentz factor.  The
exact spatial distribution of the matter traversed becomes important only when
considering radioactive isotopes, since the distance traversed in a time $T$ is
$\rho\beta c T$.

Abundances of stable secondary nuclei (produced during propagation) indicate
the average amount of interstellar medium traversed during propagation before
escape. Because secondary radioactive cosmic-ray species will be created and
decay only during transport, their steady-state abundances are sensitive to the
confinement time if their mean lifetimes are comparable or shorter than this
time.

At low energies around 1~GeV/n the abundance of individual elements in cosmic
rays has been measured. A compilation of data from various experiments is
presented in \fref{simpson}. The data are compared to the abundance in the
solar system. Overall, a good agreement between cosmic rays and solar system
matter can be stated. However, there are distinguished differences as well,
giving hints to the acceleration and propagation processes of cosmic rays.

The elements lithium, beryllium, and boron, the elements below iron ($Z=26$),
and the elements below the lead group ($Z=82$) are significantly more abundant
in cosmic rays as compared to the solar system composition (\fref{simpson}).
This is attributed to the fact, that these elements are produced in spallation
processes of heavier elements (fourth term in \eref{transporteq}) during the
propagation through the Galaxy.

\begin{figure}[t] \centering
 \epsfig{file=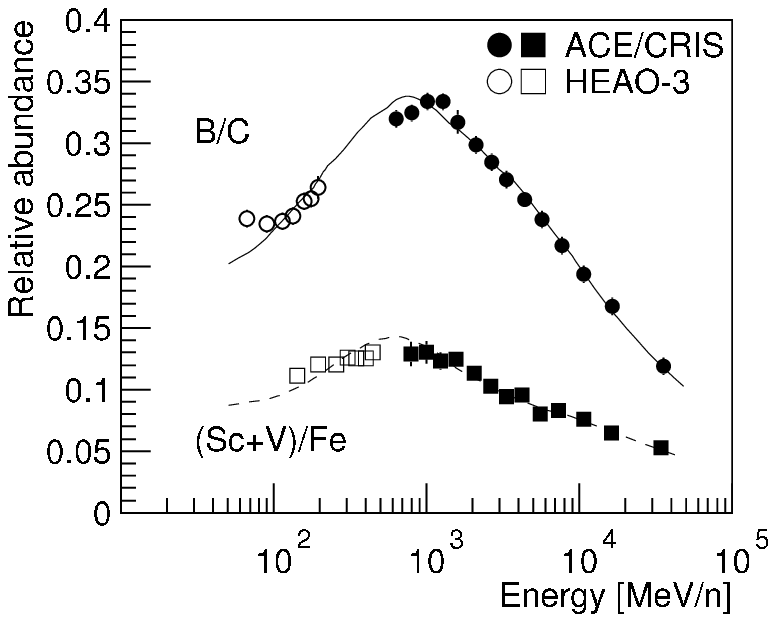,width=\columnwidth}
 \caption{Abundance ratio of boron to carbon and scandium + vanadium to iron in
	  cosmic rays as function of energy as measured by the ACE/CRIS
	  \citep{cris-time} and HEAO-3 \citep{heao3bc} experiments.  The curves
	  correspond to a Leaky Box model \eref{leakyboxfun}.}
 \label{bc}	  
\end{figure}

Measured ratios of secondary to primary nuclei, namely boron/carbon and
(scandium+vanadium)/iron, are shown in \fref{bc} \citep{cris-time}.  These
ratios can be used to estimate the matter traversed by cosmic rays in the
Galaxy \citep{garciamunoz}. The data can be successfully described using a
Leaky Box model, assuming the escape path length for particles with rigidity
$R$ and velocity $\beta=v/c$ as
\begin{equation} \label{leakyboxfun}
 \lambda_{esc}=\frac{26.7\beta ~\mbox{\gcm2}}
              {\left(\frac{\beta R}{1.0~{\rm GV}}\right)^{\delta} +
               \left(\frac{\beta R}{1.4~{\rm GV}}\right)^{-1.4}} \quad ,
\end{equation}
with $\delta=0.58$ \citep{cris-time}. The lines in \fref{bc} correspond to
this function with a path length at 2~GeV/n around 13~\gcm2 decreasing to
$\approx2$~\gcm2 at 100~GeV/n.

However, at high energies a path length according to \eref{leakyboxfun}
decreases as $\lambda_{esc}\propto E^{-\delta}$ and reaches extremely small
values. They should result in large anisotropies of the arrival direction of
cosmic rays, not observed by experiments \citep{ptuskinaniso,prop}.  To sustain
a suitable path length at high energies a residual path length model has been
proposed, assuming the relation $\lambda_{rp} = [ 6.0 \cdot \left(R/10~{\rm
GV}\right)^{-0.6} + 0.013 ]$~\gcm2 for the escape path length \citep{swordy}.
Recent measurements of the TRACER experiment yield an upper limit for the
constant term of 0.15~\gcm2 \citep{tracer05}.  

The spallation processes during the cosmic-ray propagation influence also the
shape of the spectra. Usually, it is assumed, that the energy spectra of all
elements have the same slope, i.e. the same spectral index at the source.
Taking the energy dependence of the spallation cross sections and the
pathlength of the particles in the Galaxy into account, it is found that at
Earth the spectra of heavy nuclei should be flatter as compared to light
elements \citep{prop}.  For example, the spectral index for iron nuclei should
be about 0.13 smaller than the corresponding value for protons.  Indeed, direct
measurements indicate that the spectra of heavy elements are flatter as
compared to light elements \citep{pg}, e.g. the values for protons
$\gamma_p=2.71$ and iron $\gamma_{Fe}=2.59$ differ as expected. This effect
could be of importance for the propagation of ultra-heavy nuclei and their
possible contribution to the all-particle spectrum to explain the second knee
around 400~PeV \citep{prop}.

During the propagation also radioactive secondary nuclei are produced (fifth
term in \eref{transporteq}). Their abundance can be used to estimate the
spatial distribution of the matter traversed or the cosmic-ray confinement time
in the Galaxy if the half-life time is comparable to the confinement time
\citep{garciamunoz75,garciamunoz88}.  Measuring the abundance of the isotopes
$^{10}$Be ($\tau_{1/2}=1.51\cdot10^6$~yr), $^{26}$Al
($\tau_{1/2}=8.73\cdot10^5$~yr), $^{36}$Cl ($\tau_{1/2}=3.07\cdot10^5$~yr), and
$^{54}$Mg ($\tau_{1/2}=(6.3\pm1.7)\cdot10^5$~yr) with the ACRE/CRIS experiment,
the propagation time in the Galaxy has been determined to be
$\tau_{esc}=(15.6\pm1.6)\cdot10^6$~yr \citep{cris-time}.

In the Leaky Box model, the product $\rho_{ISM}\tau_{esc}$ is proportional to
the escape path length $\lambda_{esc}=\tau_{esc}\rho_{ISM}\beta c$. Knowing
$\lambda_{esc}$ and $\tau_{esc}$, thus allows to determine the average density
of the interstellar matter.  Measurements of the CRIS experiment yield an
average hydrogen number density in the interstellar matter
$\rho_{ISM}=0.34\pm0.04$ H atoms/cm$^3$. A comparison to the average density in
the galactic disc $\rho_{GD}\approx1$/cm$^3$ indicates that cosmic rays spend
most of their propagation time outside the galactic disc, diffusing in the
galactic halo.
The height of the diffusion region into the halo has been estimated by
\cite{simon-height} with measurements of the $^{10}$Be/$^9$Be-ratio by the
ISOMAX experiment to be a few ($\approx1-4$~kpc).  Direct evidence of cosmic
rays propagating in the galactic halo is obtained from the observation of the
diffuse $\gamma$-ray background, extending well above the disc, by the EGRET
experiment \citep{egret}.  The $\gamma$-ray energy spectrum exhibits a
structure in the GeV region, which is interpreted as indication for the
interaction of propagating cosmic rays with interstellar matter
(CR+ISM$\rightarrow \pi^0 \rightarrow \gamma\gamma$) \citep{strong-moskalenko}.

At energies in the air shower domain measurements of the boron-to-carbon ratio
(or other ratios of secondary to primary elements) are not feasible any more.
However, in this energy region propagation models can be constrained by
information about the arrival direction of cosmic rays.

An anisotropy is expected due to the motion of the observer (at the Earth)
relative to the cosmic-ray gas, known as the Compton Getting effect
\citep{comptongetting}. Such an anisotropy, caused by the orbital motion of the
Earth around the Sun has been observed for cosmic rays with energies of a few
TeV \citep{cutler,tibetcg}.

The Super-Kamiokande experiment investigated large-scale anisotropies in the
arrival direction of cosmic rays with energies around 10~TeV
\citep{kamioka-anisomoriond,kamioka-aniso}. 
A $3\sigma$ excess ("Taurus excess") is found with an amplitude of $1.04\pm0.20
\cdot 10^{-3}$ at right ascension $\alpha=75^\circ\pm7^\circ$ and declination
$\delta=-5^\circ\pm9^\circ$.  On the other hand, a deficit ("Virgo deficit") is
found with an amplitude of $-(0.94\pm0.14)\cdot10^{-3}$ at
$\alpha=205^\circ\pm7^\circ$ and $\delta=5^\circ\pm10^\circ$.  
The anisotropy observed is compatible with a Compton Getting effect caused by a
velocity of about 50~km/s relative to the rest frame. This velocity is smaller
than the rotation speed of the solar system around the galactic center
($\approx220$~km/s). This implies that the rest frame of cosmic rays
(presumably the galactic magnetic fields) is corotating with the Galaxy.

The Tibet experiment reports anisotropies in the same regions on the sky
\citep{tibetscience}. For energies below 12~TeV the large-scale anisotropies
show little dependence on energy, whereas above this energy anisotropies fade
away, consistent with measurements of the KASCADE experiment in the energy
range from 0.7 to 6~PeV \citep{kascade-aniso}.  A Compton Getting effect
caused by the orbital motion of the solar system around the galactic center
would cause an excess at $\alpha\approx315^\circ$, $\delta=40^\circ$ and a
minimum at $\alpha=135^\circ$, $\delta=-49^\circ$ with an amplitude of 0.35\%.
However, the measurements at 300~TeV yield an anisotropy amplitude of
$0.03\%\pm0.03\%$, consistent with an isotropic cosmic-ray intensity. Hence, a
galactic Compton Getting effect can be excluded with a confidence level of
about $5\sigma$. This implies, similar to the result of the Super-Kamiokande
experiment, that galactic cosmic rays corotate with the local galactic
magnetic field environment.

Upper limits of anisotropy amplitudes in the PeV region are compatible with the
anisotropies expected from diffusion models of cosmic-ray propagation in the
Galaxy, while Leaky Box models predict a too high level of anisotropy
\citep[e.g.][]{prop}. In particular, a model by \cite{candiaaniso}, based on an
approach of \cite{ptuskin}, is well compatible with the measurements
\citep{maierflorenz}.

\section{Sources and Acceleration of Cosmic Rays} \label{sourcesect}

The energy density of cosmic rays amounts to about
$\rho_{cr}\approx1$~eV/cm$^3$. This value is comparable to the energy density
of the visible star light $\rho_{sl}\approx0.3$~eV/cm$^3$, the galactic
magnetic fields $B^2/2\mu_0 \approx0.25$~eV/cm$^3$, or the microwave background
$\rho_{3K}\approx0.25$~eV/cm$^3$.  The power required to sustain a constant
cosmic-ray intensity can be estimated as $L_{cr}=\rho_{cr}
V/\tau_{esc}\approx10^{41}$~erg/s, where $\tau_{esc}$ is the residence time of
cosmic rays in a volume $V$ (the Galaxy, or the galactic halo).  With a rate of
about three supernovae per century in a typical Galaxy the energy required
could be provided by a small fraction ($\approx10\%$) of the kinetic energy
released in supernova explosions. This has been realized already by
\cite{baadezwicky}.  The actual mechanism of acceleration remained mysterious
until \cite{fermi} proposed a process that involved interaction of particles
with large-scale magnetic fields in the Galaxy.  Eventually, this lead to the
currently accepted model of cosmic-ray acceleration by a first-order Fermi
mechanism that operates in strong shock fronts which are powered by the
explosions and propagate from the supernova remnant (SNR) into the interstellar
medium \citep{axford,krymsky,bell,blanford}.

Diffusive, first-order shock acceleration works by virtue of the fact that
particles gain an amount of energy $\Delta E\propto E$ at each cycle, when a
cycle consists of a particle passing from the upstream (unshocked) region to
the downstream region and back.  At each cycle, there is a probability that the
particle is lost downstream and does not return to the shock. Higher energy
particles are those that remain longer in the vicinity of the shock and have
enough time to achieve the high energy. After a time $T$ the maximum energy
attained is $E_{max}\sim Z e \beta_s B T V_s$, where $\beta_s=V_s/c$ is the
velocity of the shock.  This results in an upper limit, assuming a minimal
diffusion length equal to the Larmor radius of a particle of charge $Ze$ in the
magnetic fields $B$ behind and ahead of the shock.  Using typical values of
type II supernovae exploding in an average interstellar medium yields
$E_{max}\approx Z \cdot100$~TeV \citep{lagagemax}.  More recent estimates give a
maximum energy up to one order of magnitude larger for some types of supernovae
$E_{max}\approx Z\cdot5$~PeV \citep{berezhkomax,kobayakawa,sveshnikova}. As the
maximum energy depends on the charge $Z$, heavier nuclei (with larger $Z$) can
be accelerated to higher energies. This leads to consecutive cut-offs of the
energy spectra for individual elements proportional to their charge $Z$,
starting with the proton component.

The overall similarity between cosmic rays and matter in the solar system, as
already seen in \fref{simpson}, indicates that cosmic rays are "regular
matter", but accelerated to extreme energies.

\begin{figure}[t] \centering
 \epsfig{file=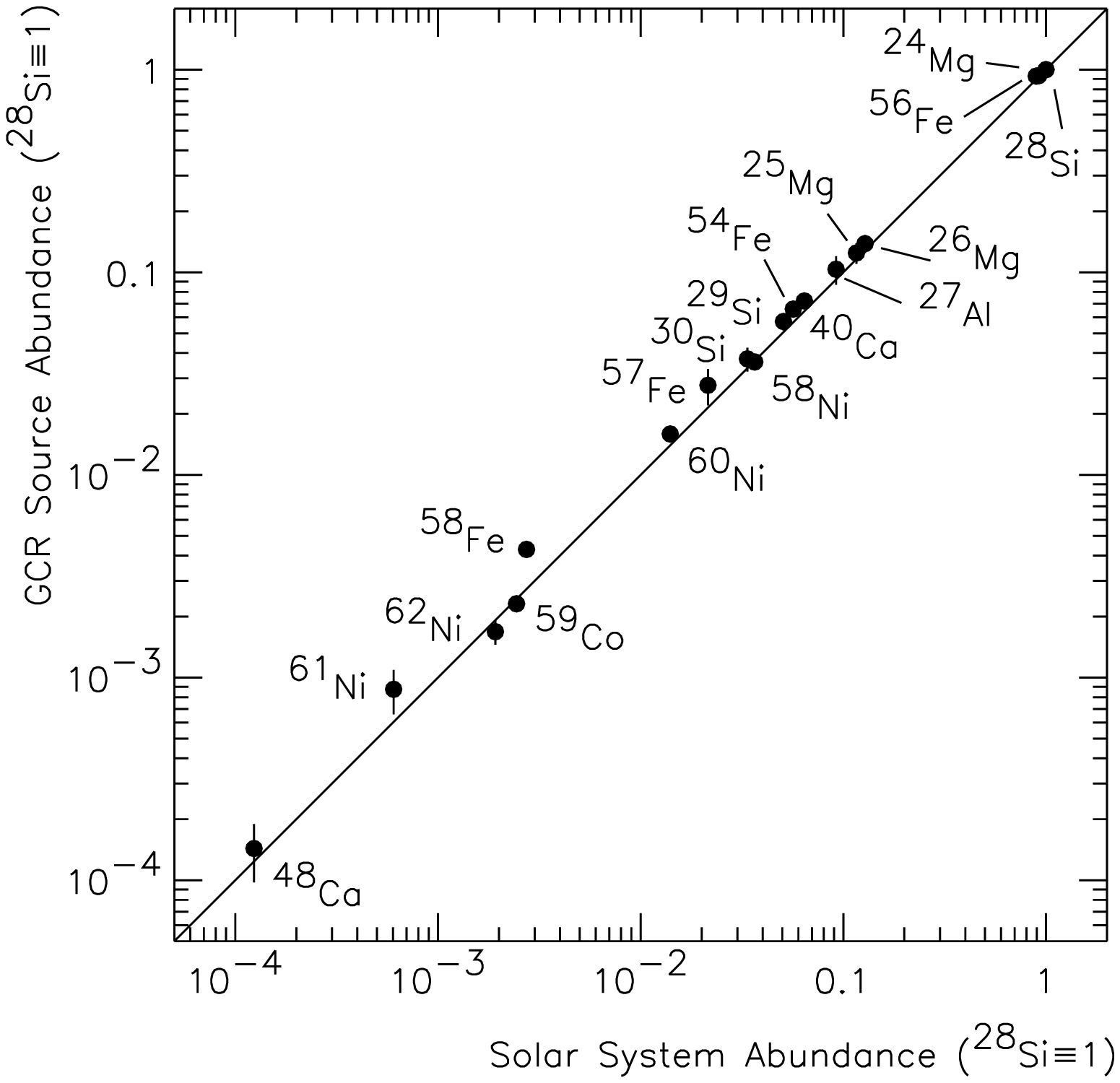,width=\columnwidth}
 \caption{Comparison of derived galactic cosmic-ray (GCR) source abundances of
	  refractory nuclides with solar-system abundances according to
	  measurements with ACE/CRIS normalized to $^{28}$Si
	  \citep{cris-abundance}.}
 \label{cris}
\end{figure}

Detailed information on the composition at the source can be obtained from
measurements of the abundance of refractory nuclei. They appear to have
undergone minimal elemental fractionation relative to one another during the
propagation process.  The derived abundance at the source is presented in
\fref{cris} versus the abundance in the solar system \citep{cris-abundance}.
The two samples exhibit an extreme similarity over a wide range. Of the 18
nuclides included in this comparison, only $^{58}$Fe is found to have an
abundance relative to $^{28}$Si that differs by more than a factor of 1.5 from
the solar-system value. When uncertainties are taken into account, all of the
other abundances are consistent with being within 20\% of the solar values.
This indicates that cosmic rays are accelerated out of a sample of well mixed
interstellar matter.

\begin{figure*}[t] \centering
 \epsfig{file=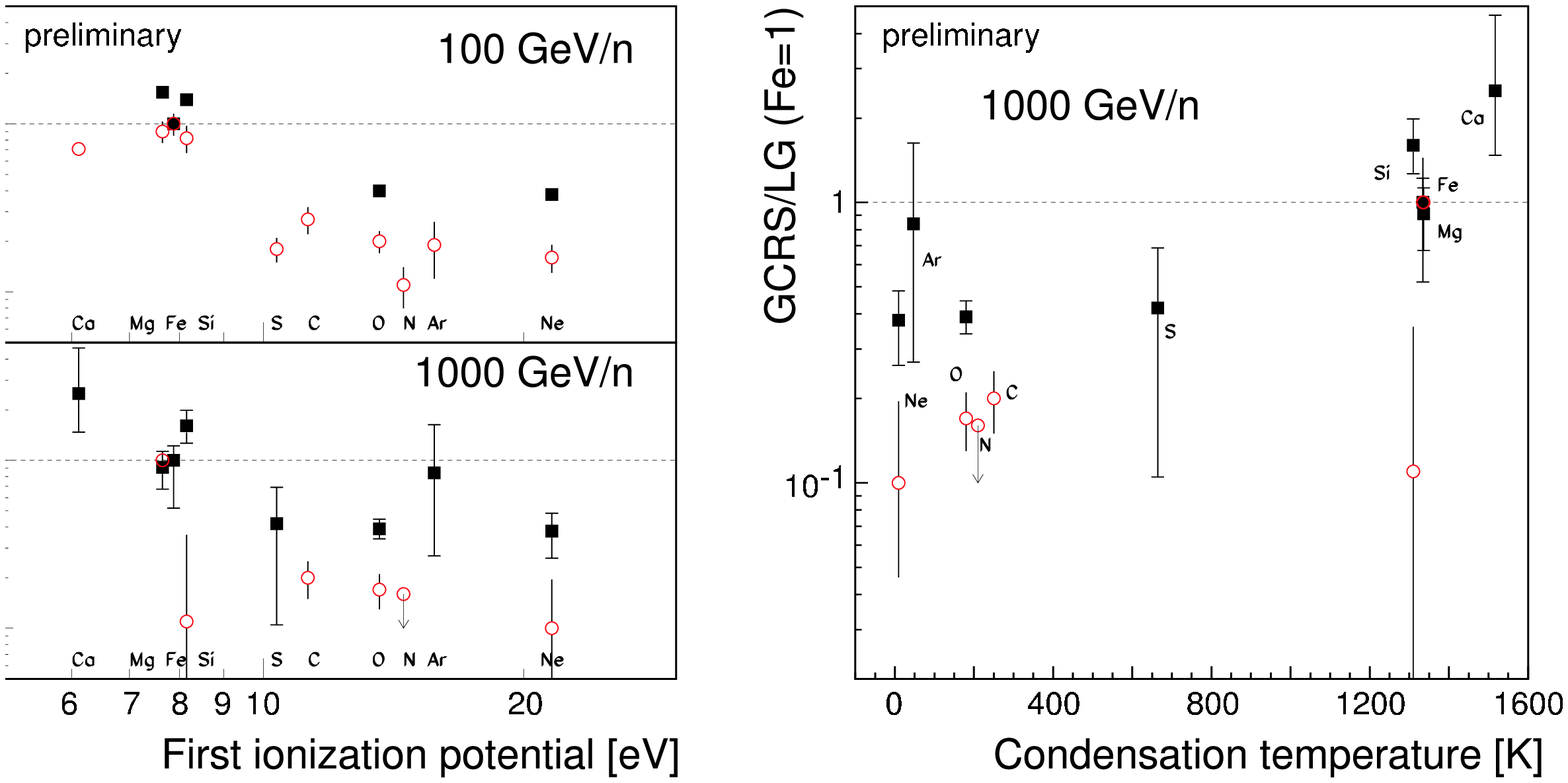,width=0.8\textwidth}
 \caption{Abundances of elements at the cosmic-ray sources relative to
	  the solar system abundances as function of the first ionization
	  potential (\lleft) and the condensation temperature (\rright) as
	  measured by the CRN and TRACER experiments \citep{tracer05}.}
 \label{tracerfip}
\end{figure*}

The time between synthesis of cosmic-ray material and its acceleration to high
energies has been estimated by measurements of the abundances of $^{59}$Ni and
$^{59}$Co by the ACE/CRIS experiment \citep{crisdelay}.  These nucleides form a
parent-daughter pair in a radioactive decay which can occur by electron capture
only. This decay cannot occur once the nuclei are accelerated to high energies
and striped of their electrons. The measured abundances indicate that the decay
of $^{59}$Ni to $^{59}$Co has occurred, leading to the conclusion that a time
longer than the $7.6\cdot10^4$~yr half-life time of $^{59}$Ni elapsed before
the particles were accelerated. Such long delays indicate the acceleration of
old, stellar or interstellar material rather than fresh supernova ejecta. For
cosmic-ray source material to have the composition of supernova ejecta would
require that these ejecta not undergo significant mixing with normal
interstellar gas before $\approx10^5$~yr have elapsed.

Ratios of individual cosmic-ray elemental abundances to the corresponding solar
system abundances seem to be ordered by the first ionization potential (FIP) of
the elements. In the GeV energy range elements with a FIP below $\approx10$~eV
are about ten times more abundant in the cosmic-ray sources than are elements
with a higher FIP \citep{meyerfip}.  It has also been found that the abundance
ratios of galactic cosmic rays to the solar system values scale with the
condensation temperature $T_c$. Refractory elements ($T_c>1250$~K) are more
abundant than volatile elements ($T_c<875$~K).  A model which could perhaps
account for the FIP effect would have a source ejecting its outer envelope for
a long period with a FIP selection effect (as the sun does) followed by a
supernova shock which sweeps up and accelerates this material. It has been
proposed that the FIP effect is due to non-volatiles being accelerated as
grains, while the volatiles are accelerated as individual nuclei
\citep{meyer1998,ellison1998}.  Recent measurements of the TRACER experiment
allow to investigate these effect at higher energies, namely at 100~GeV/n and
1000~GeV/n \citep{tracer05}.  In \fref{tracerfip} the abundance at the
cosmic-ray sources relative to the solar system abundances are shown as
function of the first ionization potential and as function of condensation
temperature.  The decreasing ratio as function of FIP and the increase as
function of condensation temperature, known from lower energies, is also
pronounced at energies as high as 1000~GeV/n, as can be inferred from the
figure.

\begin{figure}[t] \centering
 \epsfig{file=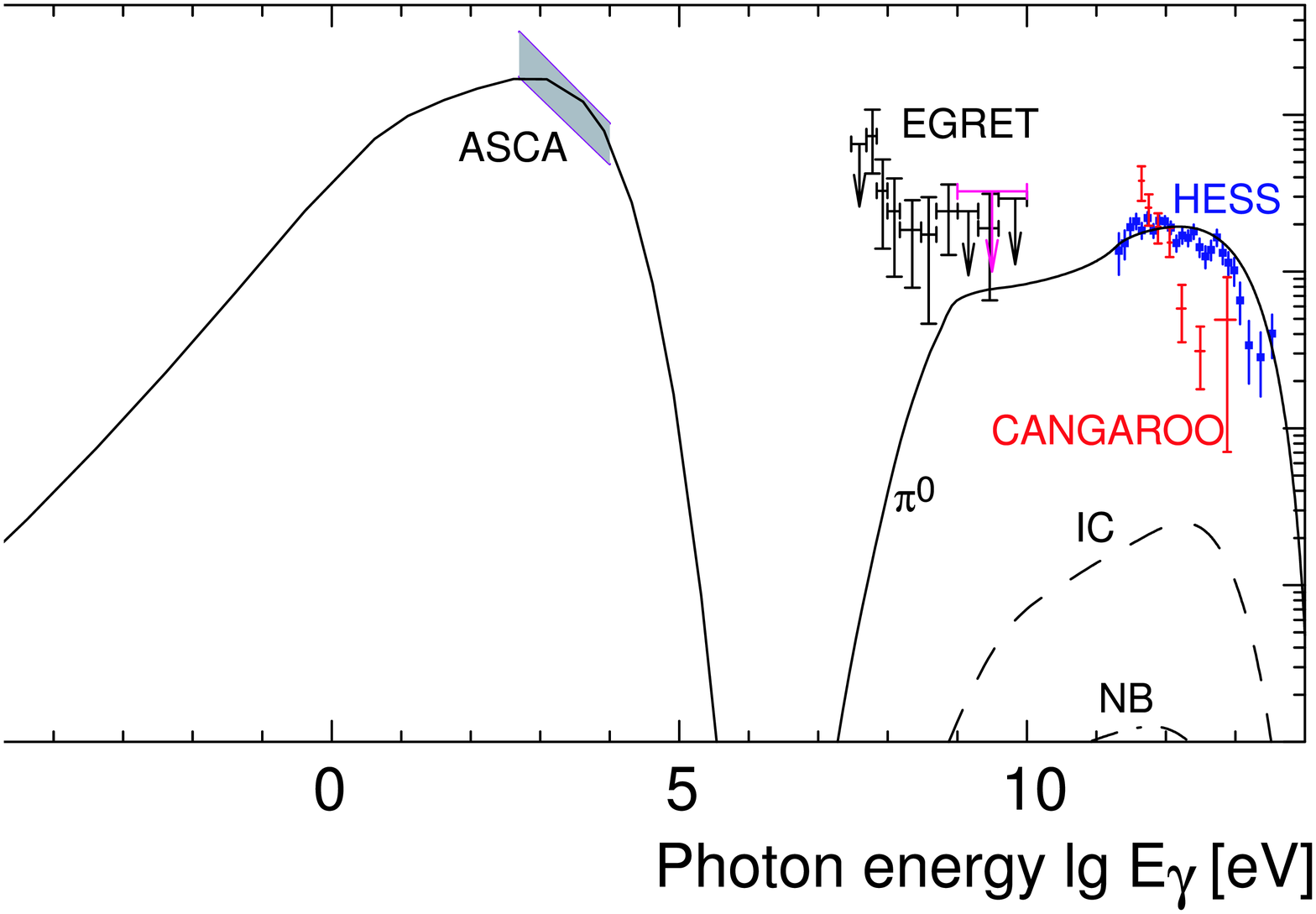,width=\columnwidth}
 \caption{Spatially integrated spectral energy distribution of the supernova
	  remnant RX J1713. The solid line above $10^7$~eV corresponds to
	  $\gamma$-ray emission from $\pi^0$-decay, whereas the dashed and
	  dash-dotted curves indicate the inverse Compton and nonthermal
	  Bremsstrahlung emissions, respectively \citep{voelkrxj1713}.}
 \label{voelk}
\end{figure}

The theory of acceleration of (hadronic) cosmic rays at shock fronts in
supernova remnants, mentioned above, is strongly supported by recent
measurements of the HESS experiment \citep{hesssnr,hessrxj1713}, observing TeV
$\gamma$-rays from the shell type supernova remnant RX J1713.7-3946,
originating from a core collapse supernova of type II/Ib. For the first time, a
SNR could be spatially resolved in $\gamma$-rays and spectra have been derived
directly at a potential cosmic-ray source. The measurements yield a spectral
index $\gamma=-2.19\pm0.09\pm0.15$ for the observed $\gamma$-ray flux.  The
photon energy spectrum of the supernova remnant RX J1713 is depicted in
\fref{voelk}.  Measurements in various energy ranges (ATCA at radio
wavelengths, ASCA x-ray, EGRET GeV $\gamma$-ray, CANGAROO and HESS TeV
$\gamma$-ray) are compared to predictions of a model by \cite{voelkrxj1713}.

The solid line below $10^6$~eV indicates synchrotron emission from electrons
ranging from radio frequencies to the x-ray regime.  The observed synchrotron
flux is used to adjust parameters in the model, which in turn, is used to
predict the flux of TeV $\gamma$-rays.  An important feature of the model is
that efficient production of nuclear cosmic rays leads to strong modifications
of the shock with large downstream magnetic fields ($B\approx100~\mu$G).  Due
to this field amplification the electrons are accelerated to lower maximum
energy and for the same radio/x-ray flux less electrons are needed.
Consequently, the inverse Compton and Bremsstrahlung fluxes are relatively low
only.  The solid line above $10^6$~eV reflects the spectra of decaying neutral
pions, generated in interactions of accelerated hadrons with material in the
vicinity of the source (hadr + ISM $\rightarrow \pi^0 \rightarrow
\gamma\gamma$).  This process is clearly dominant over electromagnetic emission
generated by inverse Compton effect and non-thermal Bremsstrahlung, as can be
inferred from the figure.  The results are compatible with a nonlinear kinetic
theory of cosmic-ray acceleration in supernova remnants and imply that this
supernova remnant is an effective source of nuclear cosmic rays, where about
10\% of the mechanical explosion energy are converted into nuclear cosmic rays
\citep{voelkrxj1713}.  Further quantitative evidence for the acceleration of
hadrons in supernova remnants is provided by measurements of the HEGRA
experiment \citep{hegra-casa} of TeV $\gamma$-rays from the SNR Cassiopeia~A
\citep{berezhko-casa} and by measurements of the HESS experiment from the SNR
"Vela Junior" (SNR RX J0852.0-4622) \citep{voelksnr}.

\section{The Knee in the Energy Spectrum} \label{kneesect}

\subsection{All-Particle Spectrum}

As discussed above, in the standard picture of galactic cosmic rays, it is
assumed that the particles gain energy in supernova remnants and propagate
diffusively through the Galaxy \citep[e.g.][]{gaissererice}. The nuclei are
accelerated up to a maximum energy, being proportional to their charge. During
propagation some particles leak out of the Galaxy and the leakage probability
is a function of the nuclear charge as well. This implies that the energy
spectra for individual elements should exhibit a break (or knee structure) at
energies being proportional to the elemental charge.

\begin{figure*}[t]\centering
  \psfig{file=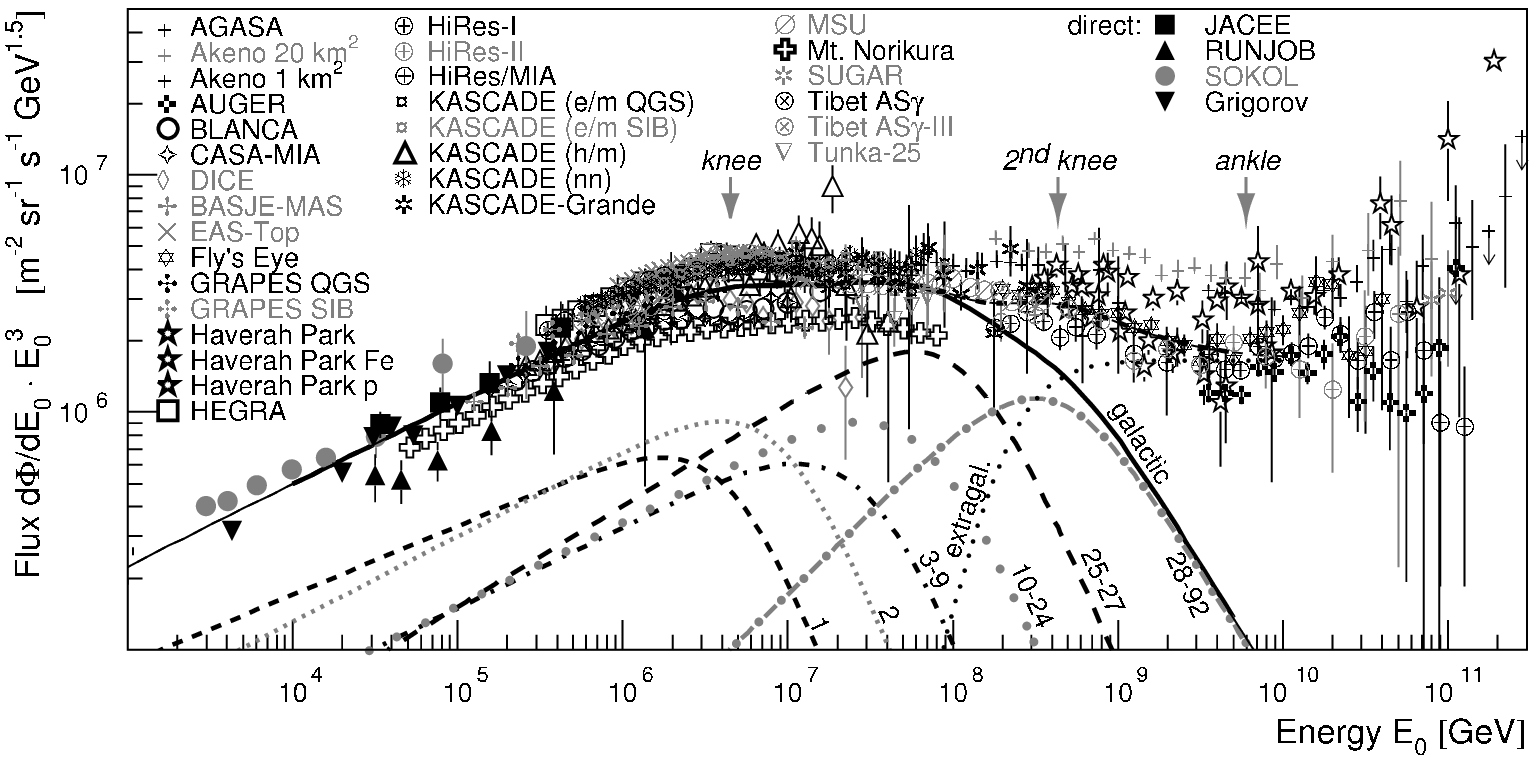,width=0.99\textwidth}%
  \caption{All-particle energy spectrum of cosmic rays, the flux is multiplied
   by $E^3$.  Results from direct measurements by
   Grigorov \etal \citep{grigorov},
   JACEE \citep{jaceefe},
   RUNJOB \citep{runjob05}, and
   SOKOL \citep{sokol}
   as well as from the air shower experiments
   AGASA \citep{agasa},
   Akeno 1~km$^2$ \citep{akeno1},
   and  20~km$^2$ \citep{akeno20},
   AUGER \citep{auger05},
   BASJE-MAS \citep{basjemas},
   BLANCA \citep{blanca},
   CASA-MIA \citep{casae},
   DICE \citep{dice},
   EAS-TOP \citep{eastope},
   Fly's Eye \citep{flyseye},
   GRAPES-3 interpreted with two hadronic interaction models \citep{grapes05},
   Haverah Park \citep{haverahpark91}
   and \citep{haverahpark03},
   HEGRA \citep{hegraairobic},
   HiRes-MIA \citep{hiresxmax},
   HiRes-I \citep{hiresi},
   HiRes-II \citep{hiresii},
   KASCADE electrons and muons interpreted with two hadronic interaction models
        \citep{ulrichapp},
        hadrons \citep{hknie}, and a neural network analysis combining
        different shower components \citep{rothnn},
   KASCADE-Grande (preliminary) \citep{kgvanburen},	
   MSU \citep{msu},
   Mt.~Norikura \citep{mtnorikura},
   SUGAR \citep{sugar},
   Tibet AS$\gamma$ \citep{tibetasg00} and
         AS$\gamma$-III \citep{tibetasg03},
   Tunka-25 \citep{tunka04}, and
   Yakutsk \citep{yakutsk5001000}.
   The lines represent spectra for elemental groups (with nuclear charge
   numbers $Z$ as indicated) according to the poly-gonato model \citep{pg}.
   The sum of all elements (galactic) and a presumably extragalactic component
   are shown as well. The dashed line indicates the average all-particle flux
   at high energies.}
  \label{espec}
\end{figure*}

The situation is sketched in \fref{espec}. The lines indicate energy spectra
for groups of elements following power laws with a break proportional to $Z$.
In the particular example the cut-off energy is assumed as
$E_k^Z=Z\cdot4.5$~PeV, according to the poly-gonato model \citep{pg} and a
possible contribution of ultra-heavy elements ($Z\ge28$) is shown as well.
Summing up the fluxes of all elements yields the all-particle flux, indicated
as "galactic" in the figure. The all-particle flux follows a power law with a
spectral index $\gamma\approx-2.7$ up to about 4.5~PeV and then exhibits a
kink, the {\sl knee} in the energy spectrum and continues with an index
$\gamma\approx-3.1$ at higher energies. This scenario agrees well with the
flux measured with detectors above the atmosphere and with air shower
experiments, shown in the figure as well.
In the measured spectrum some structures can be recognized, indicating small
changes in the spectral index $\gamma$. The most important are the {\sl knee}
at $E_k\approx4.5$~PeV, the {\sl second knee} at
$E_{2nd}\approx400$~PeV$\approx92\times E_k$, where the spectrum exhibits a
second steepening to $\gamma\approx-3.3$, and the {\sl ankle} at about 4~EeV,
above this energy the spectrum seems to flatten again to $\gamma\approx-2.7$.
To understand the origin of these structures is expected to be a key element in
understanding the origin of cosmic rays.

Various scenarios to explain the knee are proposed in the literature, for an
overview see e.g.\ \citep{origin,ecrsreview}. The most popular approaches
(maximum energy attained and leakage), just described above, are modeled with
varying details, resulting in slightly different spectra. But, also other ideas
are discussed, like the reacceleration of particles in the galactic wind, the
interaction of cosmic-ray particles with dense photon fields in the vicinity of
the sources, interactions with the neutrino background assuming massive
neutrinos, the acceleration of particles in $\gamma$-ray bursts, or
hypothetical particle physics interactions in the atmosphere, transporting
energy in unobserved channels, thus faking the knee feature.  All scenarios
result in spectra for individual elements with a break at energies being either
proportional to the nuclear charge $Z$ or to the nuclear mass $A$ which yield
certain structures in the sum spectrum. To distinguish between the different
models, measurements of the (average) mass of cosmic rays as function of energy
are required, or -- even better -- the measurement of spectra of individual
elements or at least elemental groups.

\subsection{Mean Mass}

\begin{figure*}[t] \centering
 \epsfig{file=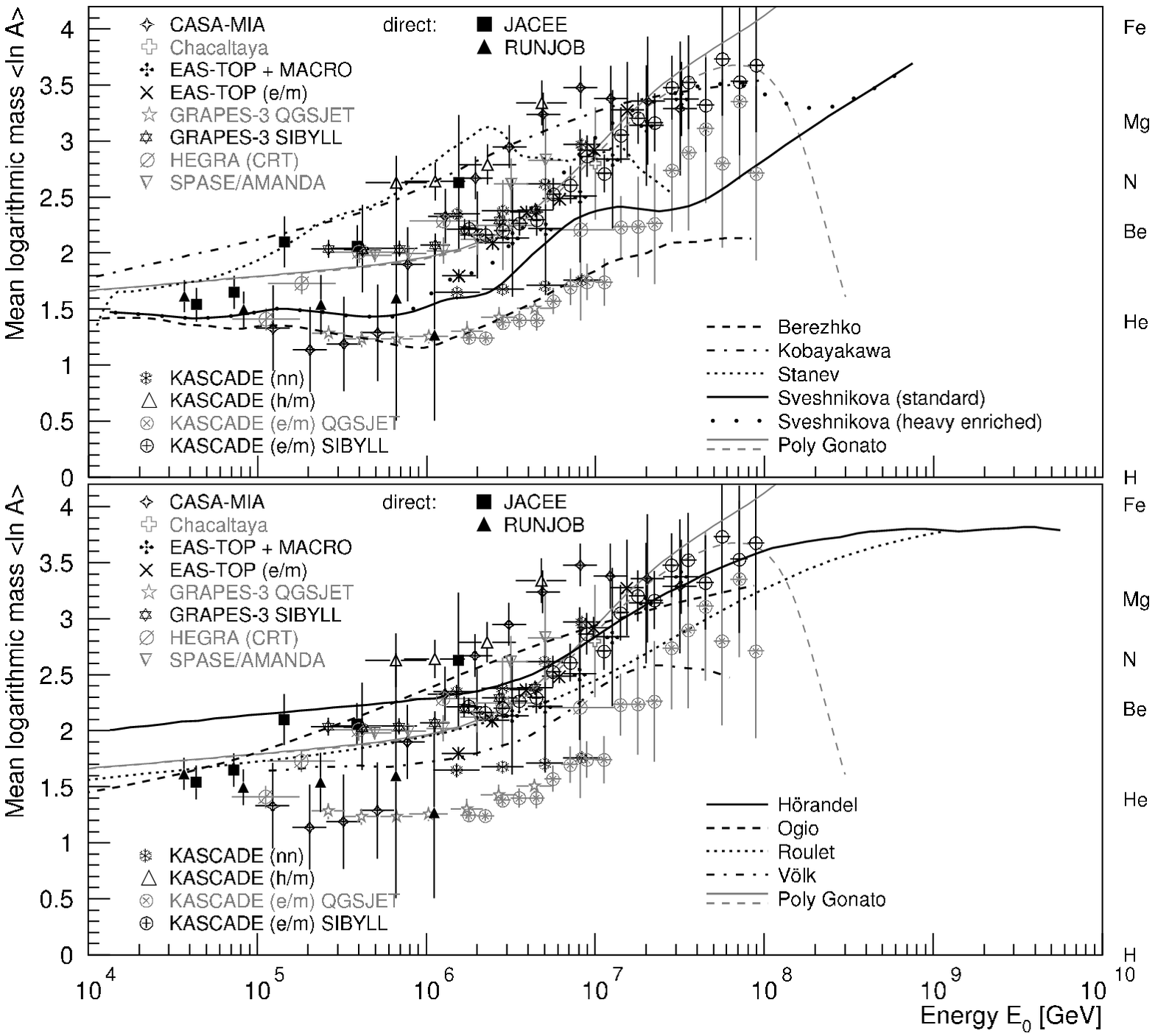,width=\textwidth}
 \caption{Mean logarithmic mass of cosmic rays derived from the measurements
           of electrons, muons, and hadrons at ground level. Results are shown
           from
           CASA-MIA \citep{casam},
           Chacaltaya \citep{chacaltaya},
           EAS-TOP electrons and GeV muons \citep{eastop-knee},
           EAS-TOP/MACRO (TeV muons) \citep{eastop-macro-lna},
	   GRAPES-3 data interpreted with two hadronic interaction models 
	   \citep{grapes05},
           HEGRA CRT \citep{hegracrt},
           KASCADE electrons and muons interpreted with two hadronic
           interaction models \citep{ulrichapp},
           hadrons and muons \citep{kascadehm}, 
           as well as an analysis combining different observables with a 
           neural network \citep{rothnn}, and
           SPASE/AMANDA \citep{spaseamandalna}.
           For comparison, results from direct measurements are shown as well
           from the JACEE \citep{jaceemasse} and RUNJOB \citep{runjob05}
           experiments.
	   For orientation, $\ln A$ for selected elements is indicated on the
	   right-hand side.
           \newline	    
           {\bf Models:}
	   The the grey solid and dashed lines indicate spectra according to
	   the \modell \citep{pg}.
           \newline   
           \TTop:
	   The lines indicate spectra for models explaining the knee due to the
	   maximum energy attained during the acceleration process according to
           \citet{sveshnikova} (\line, $\cdot~~\cdot~~\cdot$),
           \citet{berezhko} (\dashed),
           \citet{stanev} (\dotted),
           \citet{kobayakawa} (\dashdot).
           \newline   
           \BBottom:
	   The lines indicate spectra for models explaining the knee as effect
	   of leakage from the Galaxy during the propagation process according
	   to
           \citet{prop} (\line),
           \citet{ogio} (\dashed),
           \citet{roulet} (\dotted), as well as
           \citet{voelk} (\dashdot).
           }
 \label{masse}
\end{figure*}

Frequently, the ratio of the number of electrons and muons is used to determine
the mass composition, see \eref{emratioeq}.  Muons with an energy of several
100~MeV to 1~GeV are used by the experiments CASA-MIA \citep{casam}, EAS-TOP
\citep{eastop-knee}, GRAPES-3 \citep{grapes05}, or KASCADE.  To study
systematic effects two hadronic interaction models are used to interpret the
data measured with GRAPES-3 and KASCADE \citep{ulrichapp}.  High energy muons
detected deep below rock or antarctic ice are utilized by the EAS-TOP/MACRO
\citep{eastop-macro-lna} and SPASE/AMANDA \citep{spaseamandalna} experiments.
Also, the correlation between the hadronic and muonic shower components has
been investigated, e.g. by the KASCADE experiment \citep{kascadehm}. The
production height of muons has been reconstructed by HEGRA/CRT \citep{hegracrt}
and KASCADE \citep{buettner}.

\begin{figure*}[t] \centering
 \epsfig{file=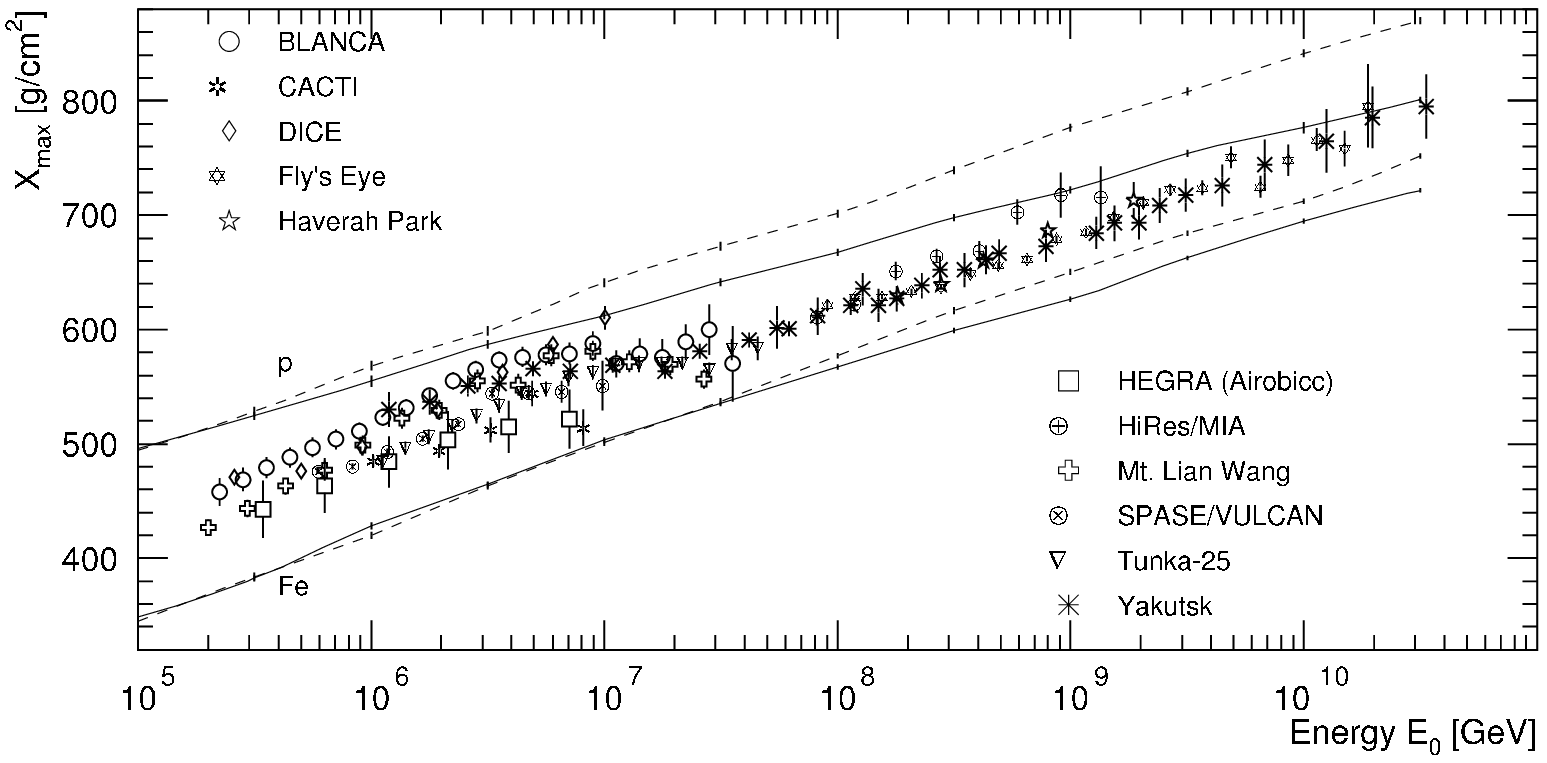,width=\textwidth}
 \caption{Average depth of the shower maximum \Xmax as function of primary
           energy as obtained by the experiments
           BLANCA \citep{blanca},
           CACTI \citep{cacti},
           DICE \citep{dice},
           Fly's Eye \citep{flyseye},
           Haverah Park \citep{haverahpark00},
           HEGRA \citep{hegraairobic},
           HiRes/MIA \citep{hires00},
           Mt. Lian Wang \citep{mtlianwang},
           SPASE/VULCAN \citep{spase99},
           Tunka-25 \citep{tunka04}, and
           Yakutsk \citep{yakutsk}. 
           The lines indicate simulations for proton and iron induced showers
           using the CORSIKA code with the hadronic interaction model QGSJET 01
           (solid line) and a version with lower cross sections and slightly
           increased elasticities (dashed line, model~3a in \citep{wq}).}
 \label{xmax}
\end{figure*}

Results from various experiments measuring electrons, muons, and hadrons at
ground level are compiled in \fref{masse}.  The same experimental data are
presented in the upper and lower graphs, where they are compared to different
models.  At low energies the values for the mean logarithmic mass are
complemented by results from direct measurements.  A clear increase of $\lnA$
as function of energy can be recognized.  However, individual experiments
exhibit systematic differences of about $\pm1$ unit.  Such fluctuations in
$\lnA$ are expected according to the simple estimate \eref{emratioeq}, assuming
that the ratio of the electromagnetic and muonic shower components can be
measured with an accuracy of the order of 16\%.  This uncertainty is a
realistic value for the resolution of air shower arrays.  Of particular
interest are the investigations of the KASCADE and GRAPES-3 experiments:
interpreting the measured data with two different models for the interactions
in the atmosphere results in a systematic difference in $\lnA$ of about 0.7 to
1.

The experimental values in \fref{masse} follow a trend predicted by the \modell
as indicated by the grey lines in the figure which implies that the increase of
the average mass with energy is compatible with subsequent breaks in the energy
spectra of individual elements, starting with the lightest species.  The
experimental values are compared to astrophysical models for the origin of the
knee in the figure. The top panel shows models which explain the knee due to
the maximum energy achieved during the acceleration process. In the lower panel
predictions from propagation models (including reacceleration during
propagation) are compiled. Details of the individual models will be discussed
below. In general, all these predictions indicate an increase of the mean mass
as function of energy very similar to the measured values.

Another technique to determine the mass of cosmic rays is to measure the
average depth of the shower maximum (\Xmax). The results of several experiments
are presented in \fref{xmax}. It depicts the measured \Xmax values as function
of energy. Two different techniques are used in the measurements, namely the
imaging technique, using telescopes to obtain a direct image of the shower
using \Cerenkov or fluorescence light and the light-integrating method, in
which the height of the shower maximum is derived from the lateral distribution
of the \Cerenkov light measured at ground level. The DICE, Fly's Eye, and HiRes
experiments use the imaging method, while all other experiments belong to the
second group.  The Haverah Park experiment uses the rise time of the shower
front to estimate \Xmax.
The data in \fref{xmax} show systematic differences of $\approx 30$~\gcm2\ at
1~PeV increasing to $\approx65$~\gcm2\ close to 10~PeV.  Some of the
experimental uncertainties may be caused by changing atmospheric conditions.
The imaging experiments measure a geometrical height which has to be converted
into an atmospheric depth.  Measuring longitudinal atmospheric profiles during
different seasons, Keilhauer \etal (2003) found that the atmospheric overburden
for a fixed geometrical height (e.g. 8~km a.s.l.) varies by at least 25~\gcm2.
The light-integrating technique is rather indirect, requiring an air shower
model to convert the observed lateral distribution into \Xmax.  Application of
different codes can explain parts of the discrepancies in \Xmax.  For example,
\cite{spase} found that using different interaction models in the MOCCA code,
viz. the original and the SIBYLL model, the systematic error amounts to $\Delta
X_{max}\approx 10$~g/cm$^2$.

The observed values are compared to predictions of air shower simulations for
primary protons and iron nuclei using the program CORSIKA \citep{corsika} with
the hadronic interaction model QGSJET~01 \citep{qgsjet} and a modified version
with lower cross sections and larger values for the elasticity of the hadronic
interactions (model~3a from \cite{wq}).  Both variations of QGSJET are
compatible with accelerator measurements within their error boundaries, for
details see \cite{wq}.  In principle, the difference between the two cases
illustrated in the figure represents an estimate of the projection of the
experimental errors from collider experiments on the average depth of the
shower maximum in air showers.  At $10^9$~GeV the difference between the two
scenarios for primary protons is about half the difference between proton and
iron induced showers. This illustrates the significance of the uncertainties of
the collider measurements for air shower observables.  The lower values for the
inelastic proton-air cross section (model~3a) are in good agreement with recent
measurements from the HiRes experiment \citep{belovisvhecri,isvhecri04wq}.  

\begin{figure}[t] \centering
 \epsfig{file=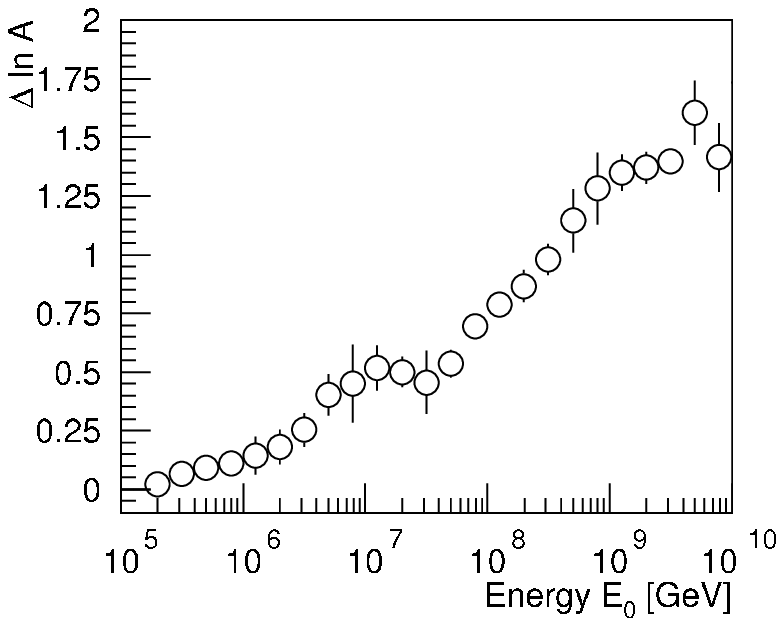,width=\columnwidth}
 \caption{Interpreting the measured \Xmax values with the two interaction
	  models, represented in \fref{xmax} by the solid and dashed lines,
	  leads to differences $\Delta\lnA$ plotted here as function of
	  energy.}
 \label{lnadiff}
\end{figure}

Knowing the average depth of the shower maximum for protons $X_{max}^p$ and
iron nuclei $X_{max}^{Fe}$ from simulations, the mean logarithmic mass can be
derived in the superposition model of air showers from the measured
$X_{max}^{meas}$ using
$\lnA=(X_{max}^{meas}-X_{max}^{p})/(X_{max}^{Fe}-X_{max}^{p}) \cdot \ln
A_{Fe}$.  This conversion requires to chose a particular interaction model.
The influence of different interaction models on the $\lnA$ values obtained is
discussed in detail elsewhere \citep{wq}.  Taking the two cases shown in
\fref{xmax} as solid and dashed lines yields differences in $\lnA$ as depicted
in \fref{lnadiff}. As expected, they grow as function of energy and exceed one
unit in $\lnA$ at energies above $10^9$~GeV.  As mentioned above, these
differences are projections of the errors of parameters derived at accelerators
on air shower measurements.  The average depth of proton showers is more
increased as the depth of iron induced cascades (see \fref{xmax}). Since in the
energy region between $10^7$ and $10^8$~GeV the measurements indicate a trend
towards a heavier composition, the influence of the modifications on the mean
mass are slightly smaller in this region, resulting in the dip in
\fref{lnadiff}.

\begin{figure*}[t] \centering
 \epsfig{file=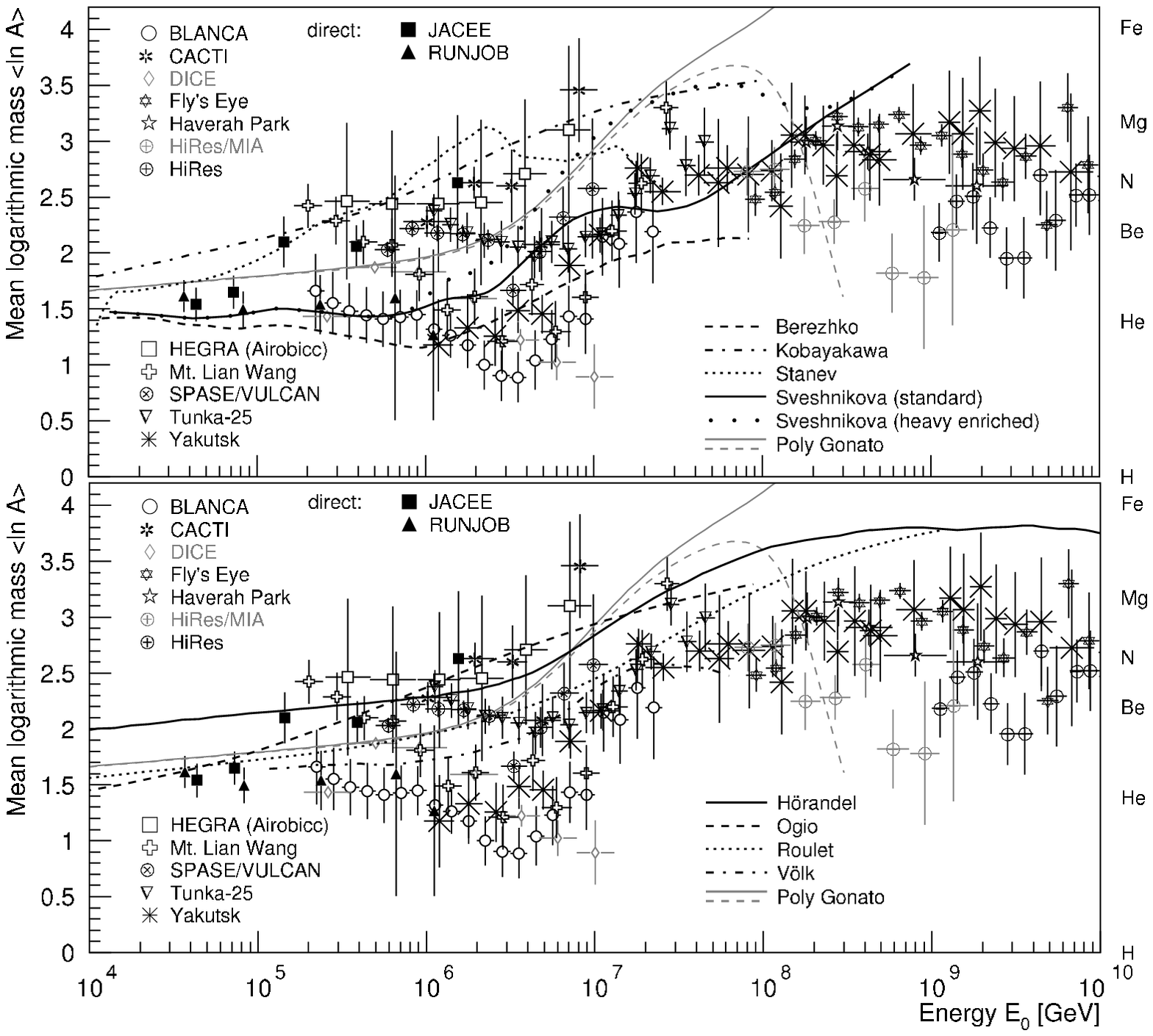,width=\textwidth}
 \caption{Mean logarithmic mass of cosmic rays derived from the average
	   depth of the shower maximum, see \fref{xmax}.  As hadronic
	   interaction model used to interpret the measurements serves a
	   modified version of QGSJET~01 with lower cross sections and a
	   slightly increased elasticity (model~3a \citep{wq}).  For
	   experimental references, see caption of \fref{xmax}. For comparison,
	   results from direct measurements are shown as well from the JACEE
	   \citep{jaceemasse} and RUNJOB \citep{runjob05} experiments.
           \newline	    
           {\bf Models:}
	   The the grey solid and dashed lines indicate spectra according to
	   the \modell \citep{pg}.
           \newline   
           \TTop:
	   The lines indicate spectra for models explaining the knee due to the
	   maximum energy attained during the acceleration process according to
           \citet{sveshnikova} (\line, $\cdot~~\cdot~~\cdot$),
           \citet{berezhko} (\dashed),
           \citet{stanev} (\dotted),
           \citet{kobayakawa} (\dashdot).
           \newline   
           \BBottom:
	   The lines indicate spectra for models explaining the knee as effect
	   of leakage from the Galaxy during the propagation process according
	   to
           \citet{prop} (\line),
           \citet{ogio} (\dashed),
           \citet{roulet} (\dotted), as well as
           \citet{voelk} (\dashdot).
	  } 
  \label{xmaxlna}          
\end{figure*}

Applying the modification of QGSJET~01 (model~3a) to the data shown in
\fref{xmax} yields $\lnA$ values plotted in \fref{xmaxlna} as function of
energy. According to \eref{xmaxeq}, an uncertainty of about 1~\xn in the
determination of \Xmax yields differences in $\lnA$ of about 1 unit.  Such a
scatter of the experimental values is visible in \fref{xmaxlna}.

Similar to the results shown in \fref{masse} also the observations of \Xmax
yield an increase of the average mass as function of energy.  The same
experimental results are shown in the upper and lower panel.  The data are
compared to astrophysical models proposed to explain the knee due to the
maximum energy attained during acceleration (upper panel) and due to leakage
from the Galaxy (lower panel).  The general trend of the data is seen as well
in the models shown.  This conclusions, of course, depends on the interaction
model used to interpret the measurements.

\subsection{Spectra for Individual Elements}

\begin{figure*}[t] \centering
 \epsfig{file=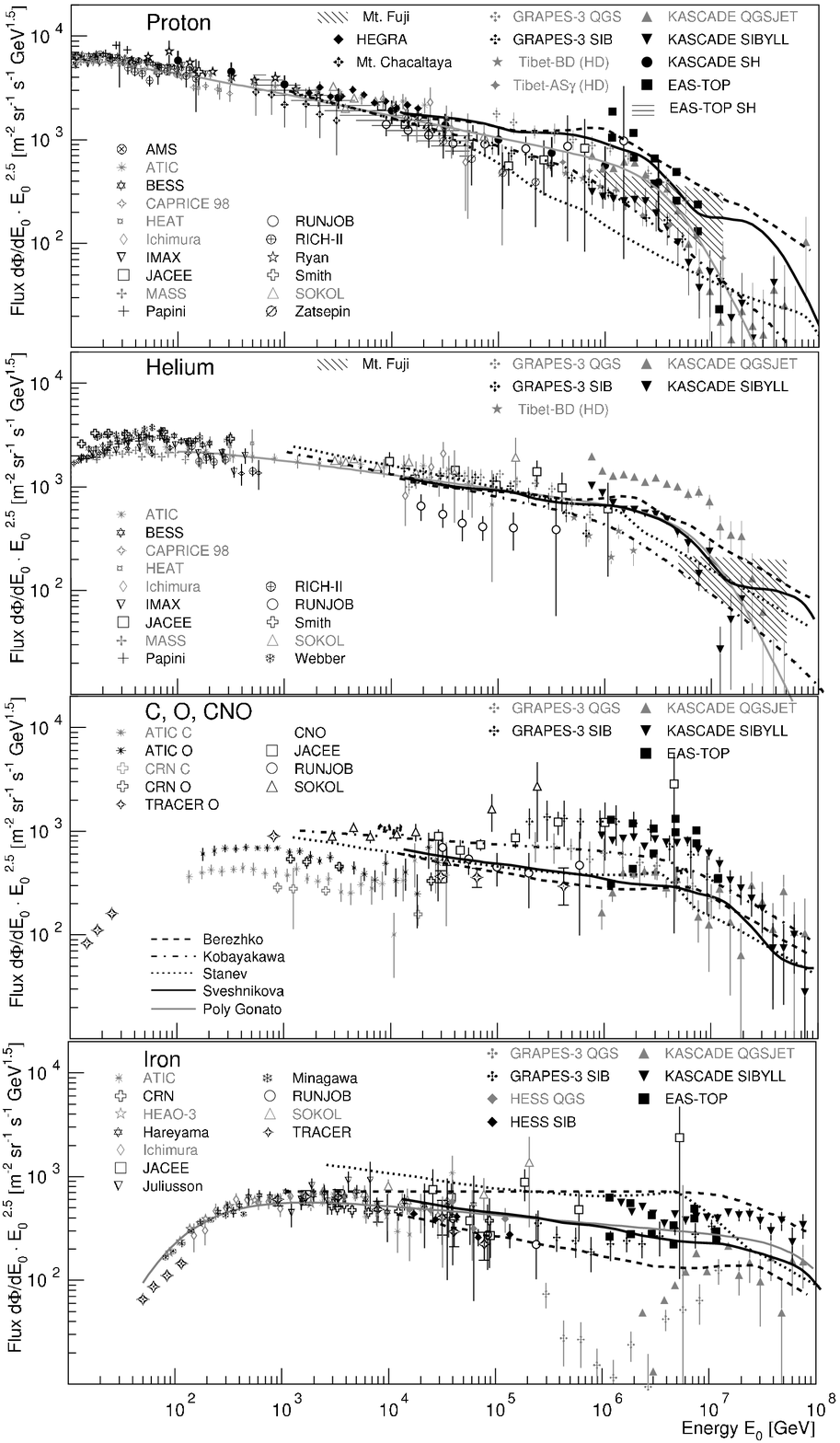,width=0.72\textwidth}
 \caption{Cosmic-ray energy spectra, caption 2 pages below.}
 \label{esnr}
\end{figure*}

\begin{figure*}[t] \centering
 \epsfig{file=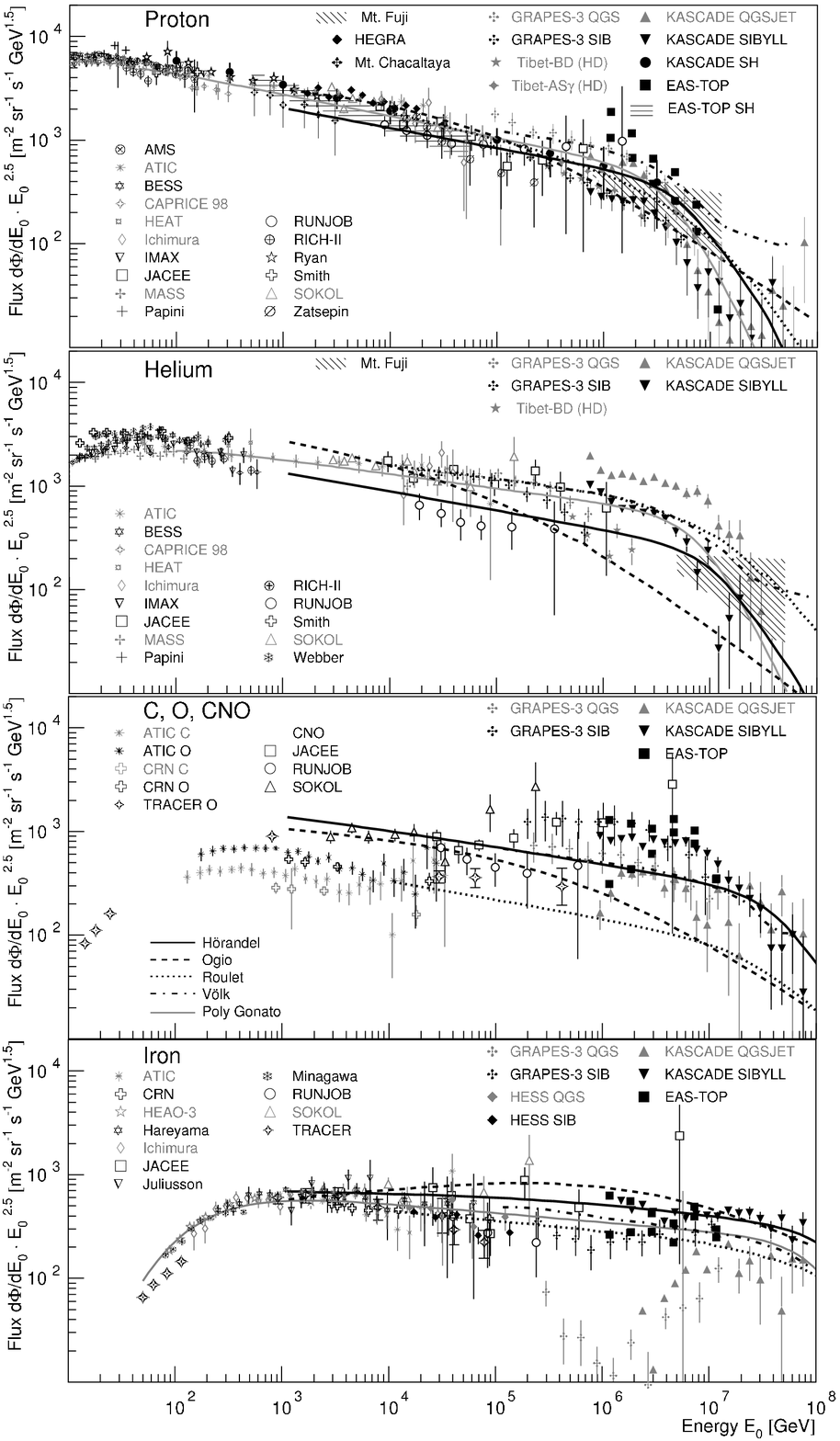,width=0.72\textwidth}
 \caption{Cosmic-ray energy spectra, caption see next page.}
 \label{eprop}
\end{figure*}

A significant step forward in understanding the origin of cosmic rays are
measurements of energy spectra for individual elements or at least groups of
elements. Up to about a PeV direct measurements have been performed with
instruments above the atmosphere \citep[e.g.][]{wiebel}. As examples,
results for primary protons, helium, the CNO group, and iron nuclei are
compiled in Figs.~\ref{esnr} and \ref{eprop}.  
The same experimental data are shown in both figures. They are compared to
different models proposed to explain the knee in the energy spectrum as
discussed in detail below.

\begin{figure*} 
 {\fref{esnr} and \ref{eprop}: Cosmic-ray energy spectra for four groups of
  elements, from top to bottom: protons, helium, CNO group, and iron group.
  \newline
   {\bf Protons:}
   Results from direct measurements above the atmosphere by 
   AMS \citep{amsp},
   ATIC \citep{atic05},
   BESS \citep{bess00},
   CAPRICE \citep{caprice98},
   HEAT \citep{heat01},
   \cite{ichimura},
   IMAX \citep{imax00},
   JACEE \citep{jaceephe},
   MASS \citep{mass99},
   \cite{papini},
   RUNJOB \citep{runjob05},
   RICH-II \citep{rich2},
   \cite{ryanp},
   \cite{smith},
   SOKOL \citep{sokol},
   \cite{zatsepinp},
   and 
   fluxes obtained from indirect measurements by
   KASCADE electrons and muons for two hadronic interaction models
   \citep{ulrichapp} and single hadrons \citep{kascadesh},
   EAS-TOP (electrons and muons) \citep{eastopspec} and single hadrons
   \citep{eastopsh},
   GRAPES-3 interpreted with two hadronic interaction models \citep{grapes05},
   HEGRA \citep{hegrap},
   Mt. Chacaltaya \citep{chacaltayap},
   Mts. Fuji and Kanbala \citep{mtfujip},
   Tibet burst detector (HD) \citep{tibetbdp} and AS$\gamma$ (HD) 
   \citep{tibetasgp}.
\newline
   {\bf Helium:}
    Results from direct measurements above the atmosphere by 
   ATIC \citep{atic05},
   BESS \citep{bess00},
   CAPRICE \citep{caprice98},
   HEAT \citep{heat01},
   \cite{ichimura},
   IMAX \citep{imax00},
   JACEE \citep{jaceephe},
   MASS \citep{mass99},
   \cite{papini},
   RICH-II \citep{rich2},
   RUNJOB \citep{runjob05},
   \cite{smith},
   SOKOL \citep{sokol},
   \cite{webberhe},
   and 
   fluxes obtained from indirect measurements by
   KASCADE electrons and muons for two hadronic interaction models
   \citep{ulrichapp},
   GRAPES-3 interpreted with two hadronic interaction models \citep{grapes05},
   Mts. Fuji and Kanbala \citep{mtfujip}, and
   Tibet burst detector (HD) \citep{tibetbdp}.
\newline
   {\bf CNO group:}
   Results from direct measurements above the atmosphere by 
   ATIC (C+O) \citep{atic06},
   CRN (C+O) \citep{crn},
   TRACER (O) \citep{tracer05},
   JACEE (CNO) \citep{jaceemasse},
   RUNJOB (CNO) \citep{runjob05},
   SOKOL (CNO) \citep{sokol},
   and 
   fluxes obtained from indirect measurements by
   KASCADE electrons and muons \citep{ulrichapp},
   GRAPES-3 \citep{grapes05}, 
   the latter two give results for two hadronic interaction models,
   and
   EAS-TOP \citep{eastopspec}.
\newline
   {\bf Iron:}
   Results from direct measurements above the atmosphere by 
    ATIC \citep{atic06},
    CRN \citep{crn},
    HEAO-3 \citep{heao3},
    \cite{juliusson},
    \cite{minagawa},
    TRACER \citep{tracer05}
    (single element resolution) and
    \cite{hareyama},
    \cite{ichimura},
    JACEE \citep{jaceefe},
    RUNJOB \citep{runjob05},
    SOKOL \citep{sokol} 
    (iron group),
   as well as
   fluxes from indirect measurements (iron group) by
   EAS-TOP \citep{eastopspec},
   KASCADE electrons and muons \citep{ulrichapp}, 
   GRAPES-3 \citep{grapes05}, and
   HESS direct \Cerenkov light \citep{hessfe}.
   The latter three experiments give results according to
   interpretations with two hadronic interaction models.
\newline
   {\bf Models:}
   The the grey solid lines indicate spectra according to the \modell
   \citep{pg}.
\newline   
   \fref{esnr}:
   The lines indicate spectra for models explaining the knee due to the maximum
   energy attained during the acceleration process according to
   \citet{sveshnikova} (\line),
   \citet{berezhko} (\dashed),
   \citet{stanev} (\dotted), and
   \citet{kobayakawa} (\dashdot)
\newline   
   \fref{eprop}:
   The lines indicate spectra for models explaining the knee as effect of
   leakage from the Galaxy during the propagation process according to
   \citet{prop} (\line),
   \citet{ogio} (\dashed),
   \citet{roulet} (\dotted), as well as
   \citet{voelk} (\dashdot).
   }
\end{figure*}   

Recently, also indirect measurements of elemental groups became possible.

A special class of events, the unaccompanied hadrons were investigated by the
EAS-TOP and KASCADE experiments \citep{eastopsh,kascadesh}. Simulations reveal
that these events, where only one hadron is registered in a large calorimeter,
are sensitive to the flux of primary protons. The derived proton fluxes agree
with the results of direct measurements as can be inferred from
\ffref{esnr} and \ref{eprop}, indicating a reasonably good understanding of the
hadronic interactions in the atmosphere for energies below 1~PeV.

At higher energies a breakthrough has been achieved by the KASCADE experiment.
Measuring simultaneously the electromagnetic and muonic component of air
showers and unfolding the two dimensional shower size distributions, the energy
spectra of five elemental groups have been derived \citep{ulrichapp}.  In order
to estimate the influence of the hadronic interaction models used in the
simulations, two models, namely QGSJET~01 and SIBYLL \citep{sibyll21}, have
been applied to interpret the measurements. It turns out that the all-particle
spectra obtained agree satisfactory well within the statistical errors. For
both interpretations the flux of light elements exhibits individual knees.  The
absolute flux values differ by about a factor of two or three between the
different interpretations.
However, it is evident that the knee in the all-particle spectrum is caused by
a depression of the flux of light elements. The KASCADE results are illustrated
in Figs.~\ref{esnr} and \ref{eprop}.

In the figures also results from further air shower experiments are shown:
EAS-TOP and GRAPES-3 derived spectra from the simultaneous observation of the
electromagnetic and muonic components.
HEGRA used an imaging \Cerenkov telescope system to measure the primary proton
flux \citep{hegrap}.
Spectra for protons and helium nuclei are obtained from emulsion chambers
exposed at Mts. Fuji and Kanbala \citep{mtfujip}.
The Tibet group performs measurements with a burst detector as well as with
emulsion chambers and an air shower array \citep{tibetbdp,tibetasg03}.
The HESS \Cerenkov telescope system derived for the first time an energy
spectrum measuring direct \Cerenkov light \citep{hessfe}.  The idea is to
register the \Cerenkov light of the primary nuclei before the air shower starts
\citep{directc}. Results for iron nuclei are shown.

Despite of the statistical and systematical error of the individual
experiments, over the wide energy range depicted, an overall picture emerges:
the measurements seem to follow power laws over several decades in energy with
a break at high energies. The bends of the spectra at low energies is due to
modulation in the heliospheric magnetic fields. 

Frequently, the question arises whether the energy spectra of protons and
helium have the same spectral index. Due to spallation of nuclei during their
propagation and the dependence of the interaction cross section on $A^{2/3}$
one would expect a slightly flatter spectrum for helium nuclei as compared to
protons. Following \cite{prop} the difference should be of the order of
$\Delta\gamma\approx0.02$.  However, a fit to the experimental data yields
$\gamma_p=-2.71 \pm0.02$ and $\gamma_{He}=-2.64\pm0.02$ (grey lines in
\ffref{esnr} and \ref{eprop}) \citep{pg}, yielding a diference
$\Delta\gamma=0.07$.

The spectra shown for carbon, oxygen, and the CNO group have a threefold
structure in resolution. At the lowest energies, clean spectra could be
resolved for carbon and oxygen only (single-element resolution). At
intermediate energies, the data represent the CNO group. The "CNO" group
derived from air shower experiments contains probably more elements than just
CNO due to the limited resolution in $\ln A$.  Thus, there is a slight step in
the flux at the transition from direct to indirect measurements.

The results shown for iron are single element spectra (as obtained e.g. by the
ATIC, CRN, HEAO-3, and TRACER experiments) as well as iron group data.
It may be worth to draw special attention to the behavior of the spectra
derived with the model QGSJET from the KASCADE and GRAPES-3 data. Both yield a
depression of the flux at $10^6$~GeV. Since both experiments obtain a similar
effect there might be some anomaly in the electron muon correlation for QGSJET
around 1~PeV. 
It may be noted as well that for both experiments QGSJET favors a lighter
composition with respect to the SIBYLL results (see \fref{masse}), i.e. an
interpretation of the measured electron-muon distributions with QGSJET yields
an increased flux of protons and helium nuclei, while on the other hand, the
CNO and iron groups are found to be less abundant in comparison to SIBYLL (see
\fref{esnr}).

For oxygen and iron it should be pointed out that the spectra of the TRACER
experiment (with single-element resolution) almost reach the energy range of
air shower measurements. It is important to realize that with a next-generation
TRACER experiment overlap in energy between direct measurements with
single-element resolution and air shower arrays could be reached, which is
important for an improvement of the understanding of hadronic interactions in
the atmosphere.

The spectra according to the \modell are indicated in the figures as solid grey
lines. It can be recognized that the measured values are compatible with breaks
at energies proportional to the nuclear charge $\hat{E}_Z=Z\cdot4.5$~PeV.

For the proton component it can be realized that the spectrum exhibits a
relatively pronounced knee. The spectral index changes by
$\Delta\gamma\approx2.1$ from $\gamma_1=-2.71$ at low energies to
$\gamma_2=-4.8$ above the knee. A similarly strong knee is visible for the
helium component.
For the iron component no clear cut-off is visible yet. An improvement of the
situation is expected by the KASCADE-Grande experiment \citep{grande}, which is
expected to unravel the energy spectra for elemental groups up to about
$10^{18}$~eV.

It is of particular interest to compare the measured spectra to various
astrophysical scenarios.
Spectra for four models, describing the acceleration of cosmic rays in
supernova remnants are shown in \fref{esnr}. In these scenarios the maximum
energy attained during the acceleration process is responsible for the knee in
the energy spectrum. The models make different assumptions on the properties
of supernova remnants and, consequently, yield different spectra at Earth.

The calculations by \cite{berezhko} are based on the nonlinear kinetic theory
of cosmic-ray acceleration in supernova remnants.  The mechanical energy
released in a supernova explosion is found in the kinetic energy of the
expanding shell of ejected matter. The cosmic-ray acceleration is a very
efficient process and more than 20\% of this energy is transfered to ionized
particles. The resulting spectra at Earth are represented as dashed lines
in the figure.

A threefold origin of energetic cosmic rays is proposed by 
\cite{stanev}. In their model particles are accelerated at three
different main sites \citep{biermann}: 
1) The explosions of normal supernovae into an approximately homogeneous
interstellar medium drive blast waves which can accelerate protons to about
$10^5$ GeV. Particles are accelerated continously during the expansion of the
spherical shock-wave, with the highest particle energy reached at the end of
the Sedov phase.
2) Explosions of stars into their former stellar wind, like that of Wolf Rayet
stars, accelerate particles to higher energies.  The maximum energy attained
depends linearly on the magnetic field and maximum energies $E_{max}=9\cdot
10^7$~GeV for protons and $E_{max}=3\cdot 10^9$~GeV for iron nuclei are
reached.
3) For energies exceeding $10^8$~GeV an extragalactic component is introduced
by the authors. The hot spots of Fanaroff Riley class II radio galaxies are
assumed to produce particles with energies up to $10^{11}$~GeV.
The spectra are indicated as dotted lines in \fref{esnr}.

A slightly modified version of the diffusive acceleration of particles in
supernova remnants is considered by \cite{kobayakawa}.  Standard first order
Fermi acceleration in supernova remnants --- with the shock normal being
perpendicular to the magnetic field lines --- is extended for magnetic fields
with arbitrary angles to the velocity of the shock front. The basic idea is
that particles are accelerated to larger energies in oblique shocks as compared
to parallel shocks. The spectra obtained are plotted as dash dotted lines in
the figure.

A usual way to increase $E_{max}$ is to enlarge the magnetic field $B$, see
e.g. the approach by \cite{kobayakawa}. On the other hand, as consequence of
recent observations also the parameters of the supernova explosion itself can
be varied. This is the basic idea of \cite{sveshnikova}, to draw up a scenario
in which the maximum energy reached in SNR acceleration is in the knee-region
of the cosmic-ray spectrum ($\approx 4$~PeV), using only the standard model of
cosmic-ray acceleration and the latest data on supernovae explosions.  Based on
recent observations the distribution of explosion energies and their rates of
occurrence in the Galaxy are estimated.  The observed spectrum in the Galaxy is
obtained as sum over all different types of supernova explosions, integrated
over the distribution of explosion energies within each supernova group.  The
corresponding spectra are shown in \fref{esnr} as black solid curves. The
structures seen in the energy spectra are a consequence of different supernova
populations.

Typically, in such models the composition is normalized to direct measurements
at energies around $10^3$ to $10^4$~GeV, which explains why the models predict
very similar fluxes in this region. On the other hand, approaching the knee,
quite some differences are visible. The least pronounced change in the spectrum
is obtained in the approach by \cite{stanev}, while the most drastic change in
the spectra is obtained in the approach by \cite{sveshnikova}.  The predictions
for the mean logarithmic mass of these models are summarized in the upper
panels of \ffref{masse} and \ref{xmaxlna}.

Examples for a second group of scenarios, in which the knee is caused by
leakage of particles out of the Galaxy, are compiled in \fref{eprop}.

\cite{ogio} consider a regular magnetic field in the Galaxy following the
direction of the spiral arms. In addition, irregularities of roughly the same
strength are supposed to exist.  It is assumed that the regular and irregular
components have about the same field strength $B\approx 3$~$\mu$G and that both
decrease exponentially with a scale hight of 1~kpc. The scale length of the
irregularities is estimated to about $L_{irr}\approx 50$~pc. The spectra
obtained are illustrated in \fref{eprop} as dashed lines.

Similar to the model discussed previously, \cite{roulet} considers the drift
and diffusion of cosmic-ray particles in the regular and irregular components
of the galactic magnetic field.  Again, a three-component structure of the
magnetic field is assumed.  The regular component is aligned with the spiral
arms, reversing its directions between consecutive arms. This field (with
strength $B_0$) will cause particles with charge $Z$ to describe helical
trajectories with a Larmor radius $R_L=p/(Z e B_0)$.  Secondly, a random
component is assumed. This will lead to a random walk and diffusion along the
magnetic field direction, characterized by a diffusion coefficient
$D_\parallel\propto E^m$. The diffusion orthogonal to the regular magnetic
field is typically much slower, however, the energy dependence of $D_\perp$ is
similar to $D_\parallel$.  The third component is the antisymmetric or Hall
diffusion, which is associated with the drift of cosmic rays moving across the
regular magnetic field. The antisymmetric diffusion coefficient is $D_A\approx
r_L c /3 \propto E$.  
For the calculations a source spectrum $dQ/dE \propto E^{-\alpha_s}$, with a
constant index for all species $\alpha_s = 2.3$ is assumed.  Below the knee,
where transverse turbulent diffusion dominates, the observed spectral index
will be $\alpha \approx \alpha_s + 1/3$, while in the drift dominated region
above $E\cdot E_k$, $\alpha\approx\alpha_s+1$ is obtained.  This results in
spectra shown in \fref{eprop} as dotted curves.  The change of the slope for
individual components $\Delta\alpha\approx2/3$ is relatively soft.

A similar approach is followed by \cite{prop}. This scenario is, as the two
models discussed above, based on an idea by \cite{ptuskin}.  Solutions of a
diffusion model are combined with numerically calculated trajectories of
particles.  While in many models the composition is normalized to values
observed at Earth, in this approach the composition at the source is assumed to
be equal to the abundance of elements in the solar system weighted with
$Z^{3.2}$.  This choice is arbitrary to a certain extent, but may be motivated
by a higher efficiency in the injection or acceleration processes for nuclei
with high charge numbers.  From the calculations follows a relatively weak
dependence of the escape path length on energy $\propto E^{-0.2}$, even weaker
as the one obtained in the model discussed above \citep{roulet}. This
necessitates relatively steep spectra at the sources $\propto E^{-2.5}$.
However, taking reacceleration of particles in the Galaxy into account would
lead to flatter spectra at the source \citep{ptuskinaspen}, being again
compatible with the standard Fermi picture and the TeV-$\gamma$ ray
observations.  The spectra obtained are represented in \fref{eprop} as black
solid lines.

Reacceleration of cosmic-ray particles in the galactic wind is discussed by
\cite{voelk}.  The wind is mainly driven by cosmic rays
and hot gas generated in the disk.  It reaches supersonic speeds at about
20~kpc above the disk, and is assumed to be very extended (several 100~kpc)
before it ends in a termination shock.  Due to galactic rotation the
differences in flow speed will lead to strong internal wind compressions,
bounded by smooth cosmic-ray shocks. These shocks are assumed to reaccelerate
the most energetic particles from the disk by about two orders of magnitude in
rigidity, ensuring a continuation of the energy spectrum beyond the knee up
to the ankle. A fraction of the reaccelerated particles will return to the
disk, filling a region around the galactic plane (several tens of kpc thick)
rather uniformly and isotropically. 
A maximum energy $E_{max}\approx Z\cdot10^{17}$~eV is obtained and the authors
conclude that the knee in the all-particle spectrum cannot be the result of the
propagation process, instead it is supposed to be a feature of the source
spectrum itself. It is pointed out that it is possible to explain the
continuation of the cosmic-ray spectrum above the knee up to the ankle in a
natural way, by considering the dynamics of the interstellar medium of the
Galaxy and its selfconsistent extension into a large-scale halo by the galactic
wind. The authors conclude further that within this picture there is no way to
produce higher energy cosmic rays, their sources must be of a different nature.

Comparing the spectra of the different models shown in \fref{eprop}, it is seen
that the approach by \cite{ogio}  yields the weakest change of the spectra in
the knee region. More pronounced changes are obtained by \cite{roulet} and
\cite{prop}.
The predictions for the mean logarithmic mass of these models are summarized in
the lower panels of \ffref{masse} and \ref{xmaxlna}.
Further models, proposed to explain the knee and their predictions are
discussed elsewhere \citep{origin,ecrsreview}.

In general, it may be remarked, that the relatively strong change in the
measured spectra (at least for protons and helium) is not reproduced by the
theoretical approaches shown in \ffref{esnr} and \ref{eprop}. It seems to be
difficult to reproduced such a relatively pronounced knee as seen in the
measurements.  \cite{prop} tried to model the relatively sharp knee by a
combination of the maximum energy attained in the acceleration process and
leakage from the Galaxy. This seems to be quite promising, as can be inferred
from the figure. This, indeed, maybe a hint that the knee is caused by a
combination of two effects, namely, the maximum energy reached during SNR
acceleration and leakage from the Galaxy.

\section{Transition to Extragalactic Cosmic Rays} \label{extragsect}

All models for the origin of the knee have in common that they predict spectra
for individual elements of galactic cosmic rays which exhibit breaks
proportional to the charge or mass of each element. This ultimately leads to
the fact that above a certain energy no more particles exist.  On the other
hand, the measured all-particle flux extends up to at least $10^{20}$~eV, and
the highest-energy particles are usually being considered of extragalactic
origin.  The Larmor radius of a proton with an energy of $10^{20}$~eV in the
galactic magnetic field is $R_L\approx36$~kpc, comparable to the diameter of
the Galaxy.  This emphasizes that such high-energy particles are of
extragalactic origin.  The transition region from galactic to extragalactic
cosmic rays is of particular interest, key features are the origin of the
second knee and the ankle.

\begin{figure}[t] \centering
 \epsfig{file=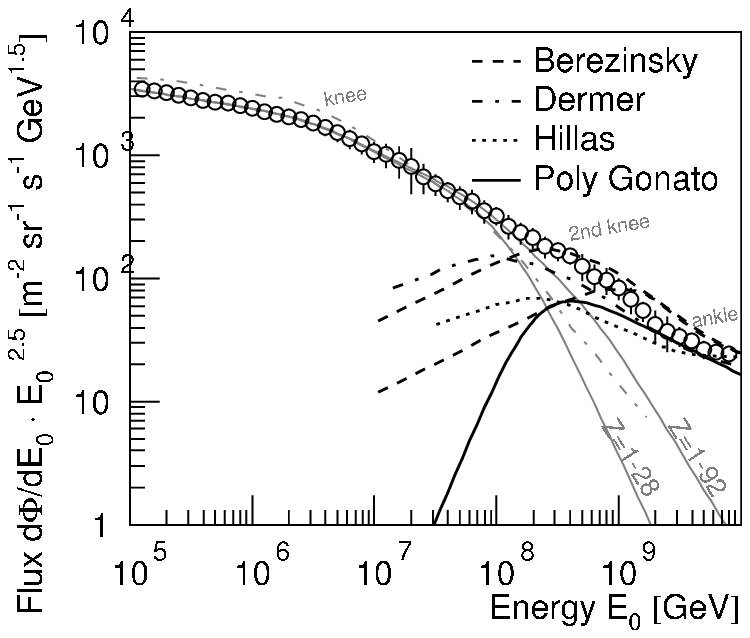,width=\columnwidth}
 \caption{Flux of cosmic rays at high energies to illustrate the transition of
          galactic to extragalactic cosmic rays.
	  The data points are average experimental values of the all-particle
	  flux \citep{pg}.
	  The black lines represent possible fluxes of extragalactic cosmic
	  rays according to models by
	  \cite{berezinskydipapp} (\dashed, two extreme cases are shown),
          \cite{hillasknee} (\dotted), 
	  \cite{dermerextrag} (\dashdot), and
	  the \modell (\line) \citep{pg}.
	  The grey lines give possible contributions of galactic cosmic rays
	  according to \cite{dermerextrag} and the \modell for elements with
	  $Z=1-28$ and $Z=1-92$.
	 } 
 \label{extragal}
\end{figure}

Different scenarios for the transition from galactic to extragalactic cosmic
rays are discussed in the literature. Some recent ideas for a possible flux of
the extragalactic component are summarized in \fref{extragal}. In these models
a possible contribution of galactic cosmic rays is estimated and subtracted
from the observed all-particle flux. The different assumptions on the behavior
of the tail of the galactic flux to highest energies yield different estimates
for an extragalactic component.

Reviewing the properties of cosmic rays accelerated in supernova remnants,
Hillas finds that a conservative estimate of the maximum energy achieved during
the acceleration in supernova remnants is not sufficient to explain the
all-particle flux up to $10^{17}$~eV.  He introduces a second (galactic)
component, accelerated by some type II supernovae, to explain the observed flux
at energies above $10^{16}$~eV \citep{hillasknee}.  He finds a contribution of
extragalactic particles (with mixed composition) as illustrated by the dotted
line in \fref{extragal}. 

In the \modell a significant contribution of ultra-heavy elements (heavier than
iron) to the all-particle flux is proposed at energies around 400~PeV
\citep{pg,prop}. In this approach the second knee is caused by the fall-off of
the heaviest elements with $Z$ up to 92. It is remarkable that the second knee
occurs at $E_{2nd}\approx92\times E_k$, the latter being the energy of the
first knee. The flux of galactic cosmic rays according to the \modell is shown
in \fref{extragal} as grey solid lines for elements with $Z=1-28$ (conservative
approach) and for all elements ($Z=1-92$). To explain the observed all-particle
flux, assuming a strong contribution of galactic ultra-heavy elements, an
extragalactic component, as illustrated by the solid line is required.
The galactic and extragalactic contributions according to this scenario are
shown as well in \fref{espec} as solid (galactic) and dotted (extragalactic)
curves.

The dip seen in the spectrum between $10^{18}$ and $10^{19}$~eV, see
\fref{espec}, is proposed to be caused by electron-positron pair production of
cosmic rays on cosmic microwave background photons $p+\gamma_{3K}\rightarrow
p+e^++e^-$ \citep{berezinskydip}.
The flux of galactic cosmic rays is extrapolated from measurements of the
KASCADE experiment. Two extreme scenarios are considered by
\cite{berezinskydipapp}, yielding extreme cases for a possible extragalactic
flux, indicated in the figure by two dashed curves.

An alternative scenario is proposed by \cite{dermerextrag}. Supposedly,
relativistic shocks from a recent ($\approx1$~Myr old) galactic gamma ray burst
($\approx1$~kpc distance to Earth) accelerate galactic cosmic rays up to
energies of the second knee. At higher energies extragalactic gamma ray bursts
are proposed to sustain the observed all-particle flux.  The corresponding
galactic and extragalactic components are indicated in the figure as dash
dotted grey and black lines.

As seen in \fref{extragal} the estimates of the various approaches are quite
similar. The extragalactic component reaches its maximum (in this $E^{2.5}$
representation) at energies around 100 to 400~PeV with the flux values
differing by about a factor of three. The differences are caused by different
estimates of the tail of the galactic flux to highest energies.

New measurements of the mass composition in the energy region of the second
knee will help to distinguish between the different models.

To reach to energies approaching $10^{18}$~eV the KASCADE experiment has been
enlarged. Covering an area of 0.5~km$^2$, 37 detector stations, containing
10~m$^2$ of plastic scintillators each, have been installed to extend the
original KASCADE set-up \citep{grande}.  Regular measurements with this new
array and the original KASCADE detectors, forming the KASCADE-Grande
experiment, are performed since summer 2003 \citep{chiavassapune}. The
objective is to reconstruct energy spectra for groups of elements up to
$10^{18}$~eV \citep{haungsaspen}.  First analyses extend the lateral
distributions of electrons and muons up to 600~m
\citep{glasstetterpune,vanburenpune}.  Based on one year of measurements,
already energies close to $10^{18}$~eV are reached.  

The Ice Cube/Ice Top experiments are installations presently under construction
at the South Pole.  Ice Cube is a cubic kilometer scale \Cerenkov detector with
the main objective to measure very high-energy neutrinos with a threshold of a
few 100~GeV \citep{icecube}. It will consist of 80 strings, each equipped with
60 digital optical modules, deployed in the Antarctic ice at depths of 1450 to
2450~m.  The ice \Cerenkov detector will also register high-energy muons
($E_\mu\ga300$~GeV) from air showers in coincidence with a surface array.  The
latter, the Ice Top experiment, is located on the snow surface above Ice Cube,
covering an area of 1~km$^2$. It consists of frozen water tanks, serving as ice
\Cerenkov detectors  to detect the electromagnetic component of air showers.
The detectors form an array of 80 detectors on a 125~m triangular grid.  It is
expected that the installations will  be completed in the austral summer
2010/11 and with this set-up air showers up to EeV energies will be
investigated.

\section{Summary and Outlook} \label{sumsect}

In the last decade substantial progress has been achieved in the understanding
of the origin of (galactic) cosmic rays. 
It has become clear that the knee in the energy spectrum at about 4~PeV is
caused by a break in the energy spectra of the light elements and the observed
knee is relatively sharp.  The mean mass of cosmic rays is found to increase as
function of energy in the knee region. 
The energy spectra for individual elements seem to follow power laws over a
wide energy range with breaks at high energies. The position of the breaks are
compatible with a rigidity dependent scenario, however, a mass dependent
behavior can not be excluded completely.
Despite of this progress, the exact astrophysical interpretation of the
measured data is limited by the present understanding of hadronic interactions
in the atmosphere.

The measurements exhibit qualitative agreement with the "standard picture".
That means the bulk of galactic cosmic rays is accelerated in shocks of
supernova remnants. The particles propagate in a diffusive process through the
Galaxy.
However, for an exact quantitative description, more detailed models are
required. 
Many contemporary models are very detailed in one aspect (e.g. acceleration),
but crude with respect to all other processes (e.g.  propagation -- and vice
versa). One should aim for integrated models, taking into account in detail the
injection, acceleration, and propagation of the particles. 
Most likely, the knee in the energy spectrum is caused by a combination of the
maximum energy reached during acceleration and leakage from the Galaxy during
propagation.  However, at present, some more exotic models cannot be excluded
completely.

Within the next decade, new measurements in the energy region around the second
knee should be able to clarify the mass composition of cosmic rays in this
region. These results are expected to contribute to the understanding of the
end of the galactic cosmic-ray spectrum and the transition to extragalactic
cosmic rays.

%\bibliographystyle{elsart-harv}
%\bibliography{cr-cospar.bib}

\end{document}